\pdfoutput=1
\documentclass[12pt, a4paper]{article}

\usepackage[height=25cm,width=16cm]{geometry}
\usepackage{wrapfig}
\usepackage{feynmp}
\usepackage{amsmath}
\usepackage{ascmac}
\usepackage{dcolumn}
\usepackage{bm,here}
\usepackage{subfig}
\usepackage{comment}
\usepackage{ifpdf}
\usepackage{slashed}
\usepackage{colortbl}
\usepackage{color}
\usepackage{comment}
\usepackage[mathscr]{eucal}
\usepackage[sort&compress, numbers, merge]{natbib}
\usepackage[symbol]{footmisc}

\usepackage{cancel}

\ifpdf
  \usepackage{graphicx} 
  \usepackage[bookmarksopen,colorlinks=true,linkcolor=bblue,citecolor=ppink,urlcolor=ppink]{hyperref}
\else 
  \usepackage[dvipdfmx]{graphicx} 
  \usepackage[dvipdfmx,bookmarksopen,colorlinks=true,linkcolor=bblue,citecolor=ppink,urlcolor=ppink]{hyperref}
\fi

\usepackage{multicol}
\definecolor{red}{rgb}{1,0,0}
\definecolor{ppink}{rgb}{0.921545,0.440586,0.687243}
\definecolor{bblue}{rgb}{0.400000,0.400000,1.000000}
\usepackage[charter]{mathdesign}
\usepackage{soul}

\newcommand{\Max}{{\rm Max}}

\begin{document}

\begin{titlepage}

\begin{flushright}
\hfill IPMU19-0087 \\
\end{flushright}

\begin{center}

\vskip 1.0cm
{\Large \bf Role of Future Lepton Colliders \\[0.5em] for Fermionic $Z$-portal Dark Matter}

\vskip 1.0cm
{\large
Dilip Kumar Ghosh$^{a,}$\footnote[1]{tpdkg@iacs.res.in},
Taisuke Katayose$^{b,}$\footnote[2]{taisuke.katayose@ipmu.jp},
Shigeki Matsumoto$^{b,}$\footnote[3]{shigeki.matsumoto@ipmu.jp}, \\[0.5em]
Ipsita Saha$^{b,}$\footnote[4]{ipsita.saha@ipmu.jp},
Satoshi Shirai$^{b,}$\footnote[5]{satoshi.shirai@ipmu.jp}
and
Tomohiko Tanabe$^{c,}$\footnote[6]{tomohiko@icepp.s.u-tokyo.ac.jp}
}

\vskip 1cm
$^{(a)}$ {\sl School of Physical Sciences, Indian Association for the Cultivation of Science,\\ 2A \& 2B Raja S.C. Mullick Road, Kolkata-700 032, India}\\[.3em]
$^{(b)}$ {\sl Kavli IPMU (WPI), UTIAS, University of Tokyo, Kashiwa, 277-8583, Japan}\\[.3em]
$^{(c)}$ {\sl ICEPP, University of Tokyo, Tokyo, 113-0033, Japan}\\[.1em]

\vskip 2.5cm
\begin{abstract}
\noindent
The fermionic $Z$-portal dark matter model suffers from severe constraints from direct detection experiments. However, a narrow parameter space around the $Z$-funnel region is beyond the present reach due to the resonance annihilation. In this paper, we provide an intriguing collider prospect for probing the $Z$-funnel dark matter mass range at the future lepton colliders including the beam polarization feature. We have done a comprehensive analysis for mono-photon signal at the colliders for such a dark matter. A realistic estimation for the 90\% C.L. constraints with the systematic beam uncertainties has also been provided.
\end{abstract}

\end{center}

\end{titlepage}

\renewcommand{\thefootnote}{\#\arabic{footnote}}
\setcounter{footnote}{0}

\tableofcontents
\newpage
\setcounter{page}{1}

\section{Introduction}
\label{sec: intro}

The discovery of the Higgs boson at the Large Hadron Collider (LHC) experiment completes the quest for the Standard Model (SM)\,\cite{Aad:2012tfa, Chatrchyan:2012ufa}. This, however, leads us to embark on the era of Higgs precision with the help of future colliders. In this regard, future lepton colliders, e.g., the International Linear Collider (ILC)\cite{Behnke:2013xla}, would be one of the best ventures to find out possible signatures of beyond the SM (BSM) physics. On the other hand, among various inadequacies of the SM, the absence of a valid Dark Matter (DM) candidate turns out to be crucial in the present situation for several reasons as follows. Most of the popular BSM theories proposed over the past few decades, such as supersymmetry, extra dimensions and composite Higgs, have been primarily advocated for solutions to the electroweak hierarchy problem. However, the absence of any sign of new particles at the LHC experiments so far, makes it difficult to have quantitative predictions for the new physics scale. Despite this, the Higgs precision data could give some indirect hint, as it is also possible that the new physics can yield a SM-like Higgs. Meanwhile, the DM abundance is precisely measured from observational evidences and the theoretical calculation of the DM abundance is usually reliable. Therefore, the DM will be the most concrete guiding principle for particle physics and hence, in the future endeavour of particle physics phenomenology, search for DM signatures at the future colliders would be one of the best bets.

Among many candidates for DM, the weakly interacting massive particles (WIMPs) are the most well-motivated. Specifically, WIMPs with a mass of the order of the electroweak scale have been studied intensively, since such a WIMP can alleviate the naturalness problem of the electroweak scale. The observed relic abundance for such a WIMP with a mass between $\mathcal{O}$(1)\,MeV\,\cite{Boehm:2002yz,Boehm:2003bt} and $\mathcal{O}(100)$\,TeV\,\cite{Griest:1989wd, Hamaguchi:2007rb, *Hamaguchi:2008rv, *Hamaguchi:2009db, Murayama:2009nj, Hambye:2009fg, Antipin:2014qva, Antipin:2015xia, Gross:2018zha, Fukuda:2018ufg} can be realized via the standard thermal freeze-out mechanism\,\cite{Bernstein:1985th,Srednicki:1988ce}. In the freeze-out scenario, the DM abundance is strongly related to the annihilation rate of WIMPs in the early Universe with the relic density being approximately proportional to the inverse of the annihilation rate. As a consequence, negligible interaction between the DM and the SM particles leads to too large DM relic abundance and conflicts with the current observation. Thus, for this mechanism to work, the DM should have sufficiently large interaction with the SM particles. One can further advantageously probe the WIMP in the collider, direct and indirect detection experiments due to the interactions.

There are, however, several exceptional cases where the interaction between the WIMP and the SM particles need not be necessarily as large\,\cite{Griest:1990kh}. For example, if the DM annihilation takes place near the pole mass of a mediator particle or in particular near the pole in the cross section, the annihilation rate is drastically enhanced. In such a case, the coupling between the WIMP and the SM particles can be adequately small and can even account for the correct DM abundance observed today. Consequently, the experimental signatures for the DM detection become weaker. Such an example is when the WIMP mass is half of the Higgs boson mass ($H$-funnel region)\,\cite{Arcadi:2019lka} or the $Z$ boson mass ($Z$-funnel region).

In this paper, we will focus on an SM gauge singlet Majorana fermion DM $(\chi)$ in the $Z$-portal scenario\,\cite{Arcadi:2014lta}, where the DM couples solely to the $Z$ boson. In this model, there are only two parameters: the DM mass $(m_{\chi})$ and the coupling constant with the $Z$ boson $(g_{\chi\chi Z})$. The observed DM abundance and the constraints from the DM spin-dependent direct detection experiments have already excluded most of the parameter space, except for the $Z$-funnel region: $m_{\chi} \simeq m_{Z}/2$. In the $Z$-funnel region, the coupling between the DM and $Z$ boson can be small to explain the DM abundance in the present universe, and accordingly the constraints of the DM direct detection experiments are still weak. We will study the prospects of probing this $Z$-funnel DM region at the future lepton colliders.

The rest of the paper is organized as follows. In Section\,\ref{sec: setup}, we briefly discuss the framework of the effective Lagrangian for the SM gauge singlet Majorana fermionic WIMP in order to introduce the $Z$-portal DM. We discuss the relic abundance condition and the current constraints from the DM (in)direct detection experiments. In Section\,\ref{sec: signal}, we study the search of the DM at lepton colliders. There are several approaches to see the DM signal at lepton colliders. We discuss the mono-photon search, the invisible $Z$ decay width measurement, and the electroweak precision measurement to see the virtual loop effects form the DM. We pay more attention to the future prospect for the mono-photon search, including the beam and detector effects, and show our result in Section\,\ref{sec: result}.

\section{The Z-portal DM model and its Cosmology}
\label{sec: setup}

\subsection{Lagrangian}
\label{subsec: setup}

We focus on a Majorana fermionic WIMP DM that is singlet under the SM gauge group. To make the WIMP stable, we impose the discrete $Z_2$ symmetry under which the WIMP is odd and all the SM particles are even. In this case, the WIMP cannot have any renormalizable couplings to the SM particles due to these symmetries. In order to introduce the interaction between the WIMP and the SM particles, we need an additional mediator particle. Physics of the WIMP strongly depends on the nature of the mediator particle. If the mediator mass is large enough than the WIMP mass and the electroweak (EW) scale, the WIMP physics can be simplified and can be described by an effective Lagrangian with the cutoff scale $\Lambda$ as long as we discuss the dynamics of the WIMP at an energy scale sufficiently lower than $\Lambda$:
\begin{eqnarray}
	{\cal L}_{\rm EFT} =
	{\cal L}_{\rm SM} +
	\frac{1}{2} \bar{\chi} (i\slashed{\partial} - m_\chi) \chi +
	{\cal L}_5 + {\cal L}_6 + {\cal L}_{\geq 7},
	\label{eq: EFT}
\end{eqnarray}
where $\chi$ is the WIMP Majorana fermion field with $m_\chi$ being its mass, and ${\cal L}_{\rm SM}$ is the SM Lagrangian. Interactions between the WIMP and SM particles are described by higher dimensional operators in ${\cal L}_{5}$, ${\cal L}_{6}$ and ${\cal L}_{\geq 7}$, which involve operators of dimension five, six, and seven or higher, respectively, which are suppressed by $\Lambda$\,\cite{Matsumoto:2014rxa}. The above effective Lagrangian is expected to be obtained by integrating out the mediator field from an appropriate original renormalizable theory, and hence $\Lambda$ represents a typical mass scale of the mediator.

We focus on the $Z$-portal DM, where the DM interacts with the SM particles through the $Z$ boson and the other interactions are suppressed\,\cite{Arcadi:2014lta}. A concrete example of the UV theory of the $Z$-portal DM is the neutralino DM in the ``blind spot"\,\cite{Cheung:2012qy}. With an appropriate choice of the neutralino mixing parameters, the dimension-five and dimension-six four-Fermi operators can be suppressed. The neutralino DM in the $Z$-funnel region can explain the correct DM abundance, while experimental tests are rather difficult \,\cite{Hamaguchi:2015rxa}. We adopt, instead, a model-independent approach to search the DM at the future lepton colliders.

As we assume that the DM is the singlet Majorana WIMP, its interaction with the $Z$ boson originates from a dimension-six operator, instead of a gauge interaction, as: 
\begin{eqnarray}
    {\cal O}_H \equiv (\bar{\chi} \gamma_\mu \gamma_5 \chi) (H^\dagger i D^{\mu} H)/2 + \rm{h.c.}, \label{eq:effective_interaction}
\end{eqnarray}
where $H$ is the SM Higgs doublet field and $D^\mu$ is the covariant derivative acting on the Higgs field\,\cite{Matsumoto:2016hbs}. By taking the unitary gauge $H = (0, v + h)^T/\sqrt{2}$ with $v \simeq 246$\,GeV being the vacuum expectation value of the Higgs doublet field and $h$ being the physical Higgs particle after the EW symmetry breaking, this operator is expanded as follows:
\begin{eqnarray}
    {\cal L}_6 \supset \frac{g_D}{\Lambda^2} {\cal O}_H
	= \frac{g_D}{4 \Lambda^2} (\bar{\chi} \gamma_\mu \gamma_5 \chi)
	\left( g_Z v^2 Z^\mu + 2 g_Z v h Z^\mu + g_Z h^2 Z^\mu   \right),
	\label{eq:O6}
\end{eqnarray}
where $Z$ is the $Z$ boson field, $g_D$ is a dimensionless coupling, and $g_Z \equiv g/\cos \theta_W$ with $g$ and $\theta_W$ being the SU(2)$_L$ gauge coupling constant and the weak mixing angle, respectively. The last three interactions on the right-hand side of the equation play negligible roles compared to the first one in the $Z$-funnel region, namely $m_\chi \sim m_Z/2$ with $m_Z$ being the mass of the $Z$ boson. Hence, we adopt the following simplified model for the $Z$-funnel WIMP DM:
\begin{eqnarray}
	{\cal L} = {\cal L}_{\rm SM} + \frac{1}{2} \bar{\chi} (i\slashed{\partial} - m_\chi) {\chi} +
	\frac{g_{\chi\chi Z}}{2}\,\bar{\chi} \slashed{Z} \gamma_5 \chi.
	\label{eq: simplified model}
\end{eqnarray}
The dimensionless coupling constant $g_{\chi \chi Z}$ is given by $g_{\chi \chi Z} = g_D g_Z v^2/(2 \Lambda^2)$, so that its range is expected to be $g_{\chi \chi Z} \lesssim 0.02\,(1\,{\rm TeV}/\Lambda)^2$. Here, we assume that the underlying model behind eq.\,(\ref{eq:O6}) is weakly coupled, namely, $g_{D} \lesssim 1$. It is also worth pointing out here that the above simplified model involves two undetermined parameters $m_\chi$ and $g_{\chi \chi Z}$, so that all the results of our discussion can be cast onto the plane spanned by the two parameters.

\subsection{Relic abundance condition}
\label{subsec: relic abundance}

\begin{figure}[t!]
	\centering
	\includegraphics[height=2.5in, angle=0]{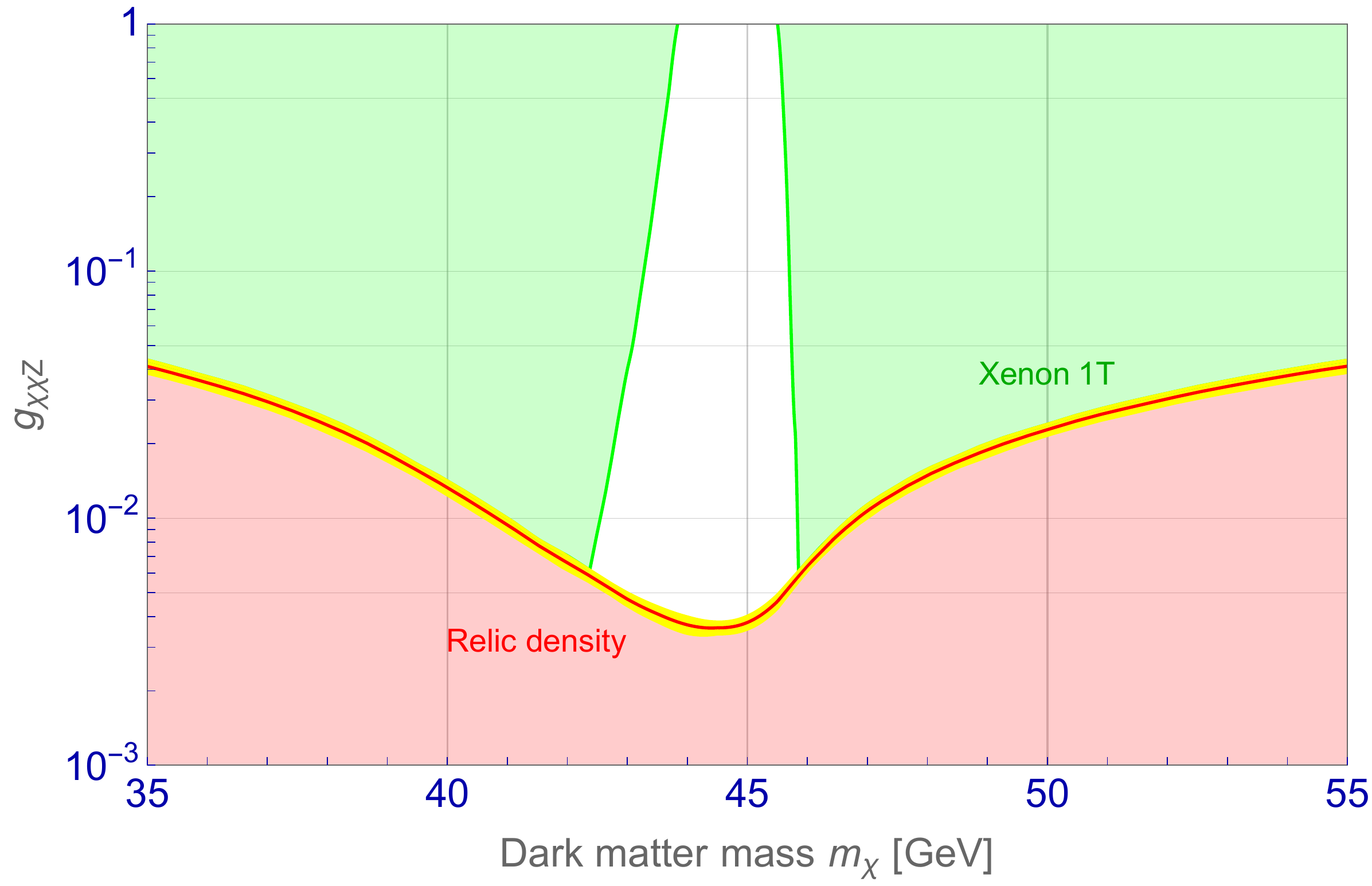}
	\caption{\small \sl The relation between the WIMP mass $m_\chi$ and the coupling constant $g_{\chi \chi Z}$ is shown as a red line, assuming that the entire DM density observed in the present universe is solely from the thermal contribution. The uncertainty from the observation and the massless degrees of freedom in the Boltzmann equation at 95\,\% C.L. are also shown as a yellow region. The shaded region below the relic density (red) line is excluded for a thermal WIMP candidate. The green shaded region is the current direct detection bound from XENON1T on this model parameter space.}
	\label{fig: relic abundance}
\end{figure}

In the early universe, the WIMP is in the thermal and chemical equilibrium with the thermal bath being composed of the SM particles, and decoupled from the bath when the temperature of the universe becomes as low as $T_f \sim m_\chi/20$\,\cite{Gondolo:1990dk}. The decoupled WIMP contributes to the DM density in the present universe, which is called the thermal contribution. The WIMP interacts with the SM particles through the $Z$ boson in the simplified model, so that the contribution can be estimated by solving the Boltzmann equation implementing this interaction. Then, if the thermal contribution explains the entire DM density observed today, it gives the relation between the DM mass $m_\chi$ and the coupling constant $g_{\chi \chi Z}$, which is shown in Fig.\,\ref{fig: relic abundance}. The uncertainty of the relation at 95\,\% C.L. from the observation\,\cite{Aghanim:2018eyx} and the massless degrees of freedom in the Boltzmann equation\,\cite{Saikawa:2018rcs} are also shown.

When the coupling constant $g_{\chi \chi Z}$ takes a value below the red line in the figure, the contribution exceeds the observed DM density. Thus, the region below the line is not attractive from the viewpoint of cosmology.\footnote{The region below the line could be consistent with the observation if, for example, the entropy of the universe is sufficiently increased by an adequate injection at the late universe. We do not consider such cases.} On the other hand, if the coupling constant takes a value above the line, the thermal contribution becomes less than the observed DM density and our WIMP model can be still viable. In such a case, the WIMP contributes to only a part of the DM density of the present universe, and the rest is composed of something different (e.g. Axion). We hence focus on the region above the line in the following discussions.



\subsection{Constraints from the direct DM detection}
\label{subsec:DD}

The direct detection of the scattering between the DM and a nucleon mediated by the $Z$ boson gives the most promising signature of the $Z$-portal DM scenario. Since the Majorana fermionic DM cannot have a vector but  an axial-vector interaction with a nucleon, the scattering takes place in a spin-dependent way and the corresponding cross section is severely constrained by the null results at various underground experiments. The spin-dependent scattering cross section between the DM and a proton\,(neutron) is given by
\begin{eqnarray}
    \sigma_{p\,(n)} = \frac{12}{\pi} \mu_{\chi p\,(n)}^2\,a_{p\,(n)}^2,
\end{eqnarray}
where $\mu_{\chi p\,(n)}$ is the reduced mass between the DM and a proton\,(neutron), while $a_{p\,(n)}$ is the scattering amplitude. A concrete expression of the amplitude is given by the formula:
\begin{eqnarray}
    a_{p\,(n)} = \frac{g_{\chi\chi Z}\,g} {8 m_Z^2 \cos\theta_W} \left(\Delta^{p\,(n)}_u - \Delta^{p\,(n)}_d - \Delta^{p\,(n)}_s \right). 
\end{eqnarray}
Here, $\Delta^{p\,(n)}_q$ is a spin nucleon parameter, and we use the default values ($\Delta^p_u = \Delta^n_d = 0.842$, $\Delta^p_d = \Delta^n_u = -0.427$, $\Delta^p_s = \Delta^n_s = -0.085$) adopted in the MicrOMEGAs code\,\cite{Airapetian:2006vy, Belanger:2018mqt}. When the DM is much heavier than the nucleon, the cross section is approximately given by
\begin{eqnarray}
    \sigma_{p\,(n)} \simeq g_{\chi \chi Z}^2\,\left[ 3.0\,(2.3) \times 10^{-37} \right]~{\rm cm}^2.
    \label{eq:unscaledDD}
\end{eqnarray}
At present, XENON1T\,\cite{Aprile:2019dbj} and PICO-60\,\cite{Amole:2019fdf} experiments give the strongest constraints on the present $Z$-portal WIMP model. In particular, when the DM mass is greater than 10\,GeV, the constraint from XENON1T experiment is stronger than the PICO-60 experiment.

When we put a constraint on WIMP DM models by the direct DM detection experiments, we often assume that all of the DM is composed of a single species of the WIMP. However, it is not always true and depends on a cosmological scenario behind it. In the cosmological scenario discussed in the previous subsection, the WIMP will contribute a fraction of the total DM density and thus the direct DM detection constraint will only be applied to the scaled scattering cross section between the DM and a nucleon (proton or neutron) as
\begin{eqnarray}
    \sigma^{\rm eff}_{p\,(n)} \equiv
    \frac{ \Omega_{\rm th}\,h^2 }
    { \Omega_{\rm obs}\,h^2}\,\sigma_{p\,(n)},
    \label{eq:effDD}
\end{eqnarray}
in the region above the red line of Fig.\,\ref{fig: relic abundance}, where $\Omega_{\rm th}\,h^2$ is the thermal contribution of the $Z$-portal WIMP to the abundance, while $\Omega_{\rm obs}\,h^2 \simeq 0.120$\,\cite{Aghanim:2018eyx} is the DM abundance observed today. Here, it is important to note that the thermal contribution $\Omega_{\rm th}\,h^2$ is inversely proportional to the annihilation cross section to a good approximation and hence $\Omega_{\rm th}\,h^2 \propto 1/g_{\chi\chi Z}^2$, while the un-scaled scattering cross section is proportional to the coupling constant squared as $\sigma_{p\,(n)} \propto g_{\chi\chi Z}^{2}$, the scaled scattering cross section $\sigma^{\rm eff}_{p\,(n)}$ is weakly dependent of the coupling $g_{\chi\chi Z}$. Such an interesting behavior of the scaled scattering cross section originates in the fact that the relic abundance and the un-scaled scattering cross section are governed by a single interaction. 

It then turns out that the present constraint from the direct DM detection excludes the region of $m_\chi \lesssim 42$--43.5\,GeV and $m_\chi \gtrsim 46$\,GeV as shown in Fig.\,\ref{fig: relic abundance}, and the $Z$-funnel region is not constrained yet. With increasing sensitivity, future LZ\,\cite{Mount:2017qzi} and PICO-500\,\cite{PICO500} experiments will be able to probe beyond $\sigma_{p(n)} \simeq 10^{-42} {\rm cm}^2$ which corresponds to $g_{\chi\chi Z} \simeq 2 \times 10^{-3}$. With these future experiments, we can probe the entire parameter region of $Z$-portal WIMP, even if you adopt the conservative effective cross section eq.\eqref{eq:effDD}. We discuss this in more details in Appendix\,\ref{app: direct detection}. 

\subsection{Constraints from the indirect DM detection}

Let us briefly discuss cosmic ray signatures from the DM annihilation at present universe. It is to be noted that being a Majorana fermion, the DM candidate in our $Z$-portal model has a significantly suppressed annihilation rate in the present Universe. The DM annihilation will take place through $s$-wave and/or $p$-wave modes. Between these two possibilities, the $s$-wave annihilation rate of the DM is suppressed by factors of $m_f/m_{\chi}$, while the $p$-wave annihilation rate is suppressed by the relative velocity of the DM, $v_{\rm rel}^2$. In the $s$-wave case, the Breit-Wigner enhancement of the DM annihilation will not work. Consequently, unlike the $p$-wave annihilation case, the velocity averaged DM annihilation rate becomes negligibly small. Quantitatively, the velocity averaged DM annihilation rate is around ($10^{-27}$\,cm$^3$/s) at least below one order of magnitude compared to the annihilation cross section limit put by the Fermi-LAT dwarf spheroidal galaxy searches\,\cite{Ackermann:2015zua}\footnote{The limit on DM annihilation should further be relaxed by an order due to the uncertainty in the J-factor.}. In passing, we would like to mention that the small propagator width effect may play a role in the $Z$-funnel mass region and the $p$-wave annihilation becomes dominant. However, due to the low relative DM velocity of ${\cal O}(10^{-3} c)$ in the present galactic objects, the $p$-wave annihilation is also strongly suppressed and even lower than the $s$-wave contribution even for $m_\chi \sim M_Z/2$. Hence, there is no significant constraint on this model from the cosmic-ray observation.

\section{WIMP signals at collider experiments}
\label{sec: signal}

We discuss how the fermionic WIMP in the $Z$-funnel region is searched for at the colliders. We first discuss the lepton collider signature. We consider the mono-photon channel associated with the WIMP pair production and figure out which center-of-mass energy and polarizations of incident electron and positron beams are desired to efficiently search for it. Next, we consider the measurement of the $Z$ boson invisible decay width for the WIMP search, whose sensitivity will be compared with that of the mono-photon channel in the following sections. We also discuss the role of the electroweak precision measurement for the WIMP. Finally, we address the present constraint on the WIMP obtained by the LEP/LHC experiments, which will be also compared with the case of the future lepton colliders.

\subsection{Mono-photon search}

\subsubsection{Signal and background processes}

\begin{figure}[t!]
	\centering
	\includegraphics[height=2.8in, angle=0]{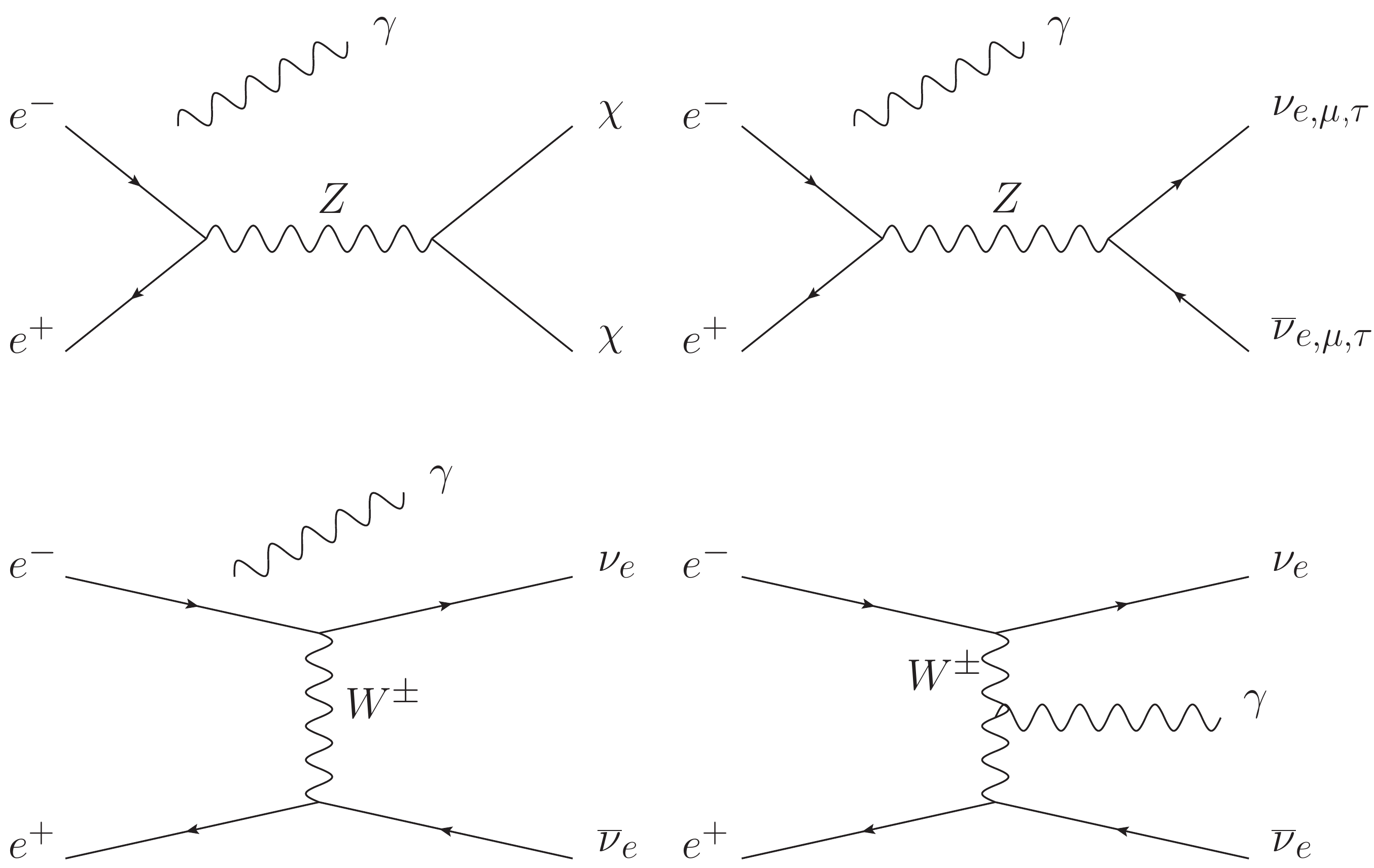}
	\vspace{0.5cm}
	\caption{\small \sl Feynman diagrams for the signal process (top-left) and irreducible background processes (others) of the mono-photon search at lepton colliders. Here, the photon line which is not directly touched onto other lines means that it can be from either electron or positron line at the initial state.}
	\label{fig: diagrams}
\end{figure}

Since DM cannot be directly captured by collider detectors, we search for it indirectly through, for instance, the observation of a recoiled SM particle against the DM pair production. Among various channels to search for the DM, the mono-photon process ($e^- e^+ \to \chi \chi \gamma$) is known to be one of the most efficient channels at the lepton colliders. The mono-photon signal in the framework of the simplified model in eq.\,(\ref{eq: simplified model}) is from Feynman diagrams shown in Fig.\,\ref{fig: diagrams} (the top-left diagram), where the photon line which is not directly touched onto other lines means that it can be from either initial electron or positron line.

On the other hand, there are several SM processes contributing to the mono-photon channel as backgrounds against the signal. One of such backgrounds is an irreducible one coming from the neutrino pair production associated with a photon ($e^- e^+ \to \nu\,\bar{\nu}\,\gamma$), whose diagrams are also shown in Fig.\,\ref{fig: diagrams}. The Bhabha scattering process with a photon emission ($e^- e^+ \to e^- e^+ \gamma$) can also be a background if both electron and positron at the final state go to the beam pipe direction. This background is, fortunately, reduced efficiently by considering only events with a large photon transverse momentum in the analysis. Other possible backgrounds come from the neutrino pair production associated with more than one photon ($e^- e^+ \to \nu\,\bar{\nu}\,\gamma{\rm s}$). These contributions can be taken into account as the effect of the initial state radiation, as will be discussed in the next section. Finally, multi photon productions from the $e^- e^+$ annihilation ($e^- e^+ \to \gamma{\rm s}$) can be backgrounds if some photons in the final state are failed to be detected, though these are not significant compared to the irreducible background under an appropriate event selection\,\cite{Bartels:2012ex}. We therefore only consider the irreducible background as the one against the mono-photon signal.

\subsubsection{Optimizing the center-of-mass energy and polarizations}
\label{subsubsec: optimization}

\begin{figure}[t!]
	\centering
	\includegraphics[height=2.5in, angle=0]{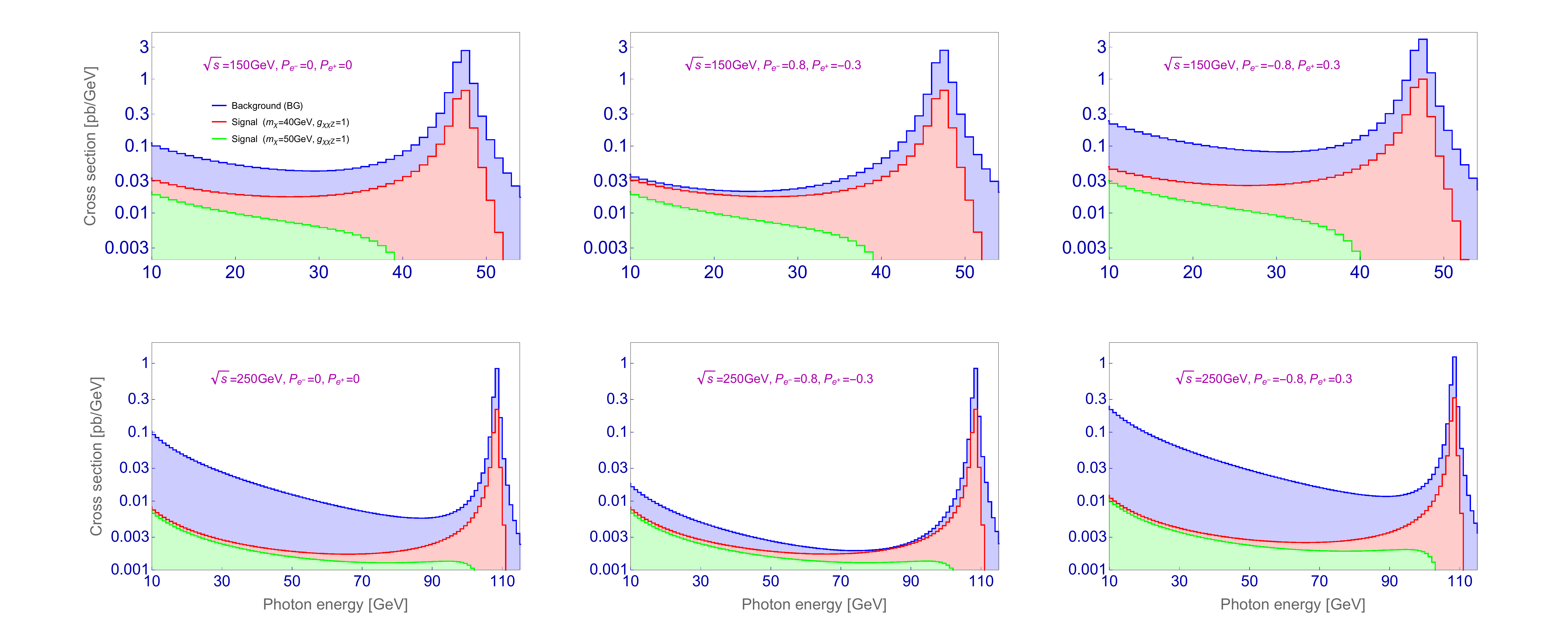}
	\caption{\small \sl The differential cross sections for the signal with $m_\chi = 40$\,GeV and the signal with $m_\chi = 50$\,GeV, and the SM background are shown with the coupling constant $g_{\chi \chi Z}$ being fixed to be one for several choices of the center-of-mass energy $\sqrt{s}$ and polarizations ($P_{e^-}$, $P_{e+}$). These peaks correspond to the on-shell production of Z boson and its decaying into neutrinos or DMs, therefore green lines ($m_{\chi}=50\,GeV>m_Z/2$) do not have any peak.}
	\label{fig: cross section}
\end{figure}

An important question here is how the signal can be distinguished from the background by observing only one photon. The photon is characterized by two quantities; its energy ($E_\gamma$) and the scattering angle ($\cos \theta_\gamma$). Hence, the quantitative question is on the differential cross sections for the signal and the background processes on the plane spanned by the two quantities. Since both the cross sections do not have a characteristic feature on the $\cos \theta_\gamma$ dependence, we focus on the $E_\gamma$ dependence of the cross sections. The cross section at each energy bin, $d \sigma/dE_\gamma$, which is obtained by integrating the differential cross sections over the range of $|\cos \theta| \leq 0.98$, is shown in Fig.\,\ref{fig: cross section}, where those of the irreducible background (BG), the signal with $m_\chi = 40$\,GeV and the signal with $m_\chi = 50$\,GeV, with the coupling $g_{\chi \chi Z}$ being fixed to be one, are depicted as blue, orange and green lines, respectively, for several choices of the center-of-mass energy ($\sqrt{s}$) and polarizations ($P_e$).\footnote{$P_{e^-} = \pm 0.8$ means 90\,\% (10\,\%) of the incident electron is right-handed and the rest 10\,\% (90\,\%) is left-handed, while $P_{e^+} = \mp 0.3$ means 65\,\% (35\,\%) of the incident positron is right-handed and the rest 35\,\% (65\,\%) is left-handed. Note that $P_{e^-} = \pm 0.8$ and $P_{e^+} = \mp 0.3$ are maximal polarizations that ILC can achieve\,\cite{Behnke:2013xla}.} The figure shows that the signal-to-background ratio depends considerably on the choice of $\sqrt{s}$ and $P_e$, so that it is important to optimize these two values in order to search for the WIMP efficiently.

\begin{figure}[t!]
	\centering
	\includegraphics[height=2.0in, angle=0]{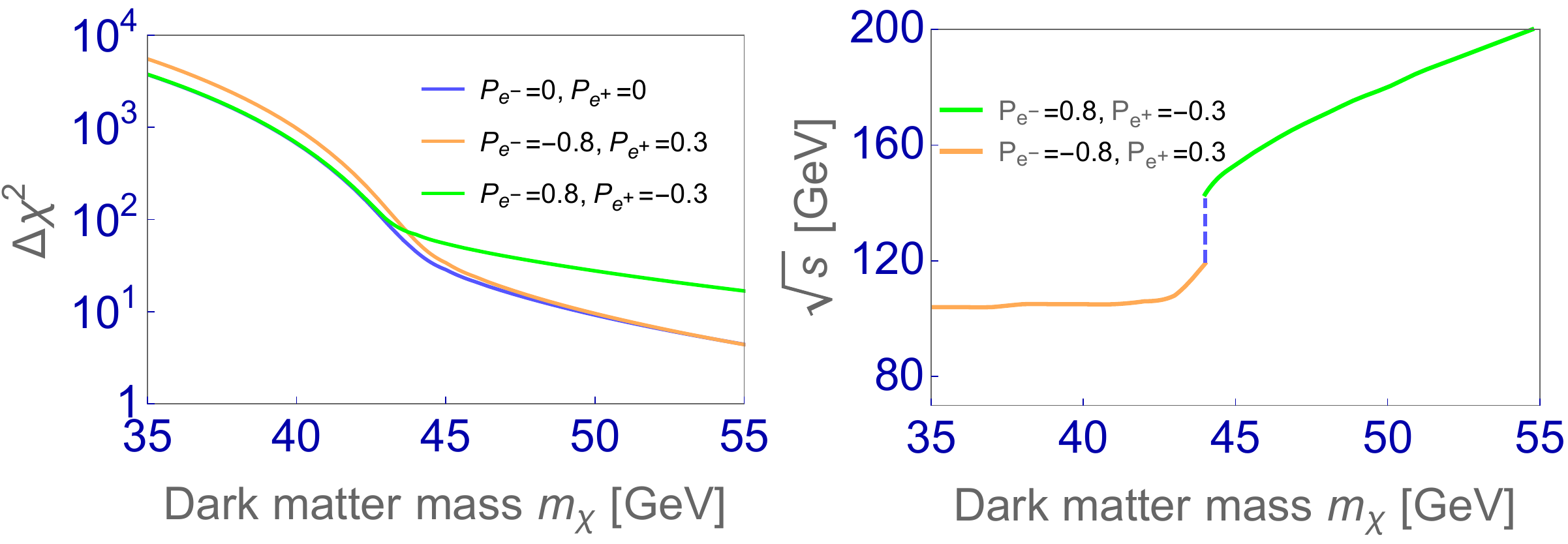}
	\caption{\small \sl {\bf (Left panel)} The value of $\Delta \chi^2$ is shown for several choices of the polarizations with the center-of-mass energy being optimized so that it gives the maximal $\Delta \chi^2$ value for each DM mass. The coupling constant and luminosity are $g_{\chi \chi Z} = 0.1$ and 2\,ab$^{-1}$, respectively. {\bf (Right panel)} The optimized center-of-mass energy and the polarizations of incident electron and positron to make the significance of the signal event against the background event maximal. See text for details.}
	\label{fig: optimization}
\end{figure}

In order to quantify this efficiency, we consider the significance of the signal event using the likelihood analysis with the so-called $\Delta \chi^2$ value defined by the following formula:
\begin{eqnarray}
	\Delta \chi^2 \equiv \sum_i \frac{ (N_i - N^{\rm BG}_i)^2 } {N^{\rm BG}_i },
	\label{eq: delta chi2}
\end{eqnarray}
where $N_i$ is the expected number of the events of the signal plus the SM background at the `$i$'-th energy bin with the bin-width of 1\,GeV. We consider the range of $E_\gamma \geq 10$ GeV and $|\cos \theta_\gamma| \leq 0.98$ to compute $N_i$s, which validates ignoring other backgrounds in our analysis as mentioned above. The expected number of the background event is denoted by $N^{\rm BG}_i$, which is computed in the same manner. In the left panel of Fig.\,\ref{fig: optimization}, the value of $\Delta \chi^2$ is shown for several choices of the polarizations, ($P_{e^-}$, $P_{e^+}$) = (0, 0), (0.8, -0.3) and (-0.8, 0.3), where the center-of-mass energy is fixed so that it gives the maximal $\Delta \chi^2$ value at each DM mass. Here, $g_{\chi \chi Z} = 0.1$ and the luminosity is fixed to be 2\,ab$^{-1}$.\footnote{Since the new physics contribution to the $Z$ boson decay width (the $Z$ decay into the DM pair) is negligible whenever $g_{\chi \chi Z}$ is small, the value of $\Delta \chi^2$ is simply proportional to $g_{\chi \chi Z}^4$ to a good approximation.}

The figure shows that the right-handed beam polarization is more efficient than the left-handed one to search for the signal when $m_\chi \gtrsim m_Z/2$, as background events originating in the weak interaction are suppressed. On the other hand, the left-handed beam polarization becomes more efficient when $m_\chi \lesssim m_Z/2$, because the same on-shell $Z$ production process in Fig.\,\ref{fig: diagrams} dominates for both signal and background, and their cross sections become maximal when the beam polarization is left-handed. Hence, We can adopt the left-handed polarization, ($P_{e^-}$, $P_{e^+}$) = (-0.8, 0.3), for $m_\chi \leq 44$\,GeV and the right-handed one, ($P_{e^-}$, $P_{e^+}$) = (0.8, -0.3), for $m_\chi \geq 44$\,GeV, with the center-of-mass energy being fixed so that $\Delta \chi^2$ becomes maximal. This setup is summarized in the right-panel of Fig.\,\ref{fig: optimization} as a function of $m_\chi$.

\subsection{Invisible $Z$ decay}
\label{subsec: Inv Z}

When the DM mass $m_\chi$ is less than $m_Z/2$, the $Z$ boson can decay into a pair of DMs in addition to ordinary decay channels into SM particles. Hence, it is possible to search for the DM by observing the invisible decay width of the $Z$ boson. In the framework of the simplified model\,(\ref{eq: simplified model}), the new physics contribution to the width is predicted as follows:
\begin{eqnarray}
	\Gamma(Z \to \chi \chi) = \frac{g_{\chi \chi Z}^2 m_Z}{24 \pi} \left(1 - \frac{4 m_\chi^2}{m_Z^2} \right)^{3/2}.
\end{eqnarray}
The SM process, namely the $Z$ boson decay into a neutrino pair, also contributes to the invisible decay width, which is proportional to the number of the neutrino flavors $N_\nu$.

On the other hand, the invisible decay width of the $Z$ boson is experimentally determined by comparing the total decay width and observable partial decay widths of the $Z$ boson. At present, the invisible decay width is observed to be $\Gamma^{(Z)}_{\rm inv} = 499.0 \pm 1.5$\,MeV at the LEP experiment\,\cite{Patrignani:2016xqp}, which is translated to a constraint on the new physics contribution to the invisible decay width as $\Gamma(Z \to \chi \chi) \leq 1.5$\,MeV, or in other words, $\Delta N_\nu \leq 8 \times 10^{-3}$ with $\Delta N_\nu$ being the new physics contribution to the number of the neutrino flavors\,\cite{Carena:2003aj}.

The precision of the width measurement can be improved at the future lepton colliders. For instance, the Circular Electron Positron Collider (CEPC) experiment has a potential to determine the width very precisely due to its circular nature; it will accumulate more data of the $Z$ boson than ILC. Assuming that the expected sensitivity of CEPC is the same as the TLEP experiment\,\cite{Gomez-Ceballos:2013zzn}, the constraint will be updated to be $\Gamma(Z \to \chi \chi) \leq 0.01$\,MeV (or in other words, $\Delta N_\nu \leq 6 \times 10^{-5}$), if no new physics contribution is observed, where only the statistical uncertainty is involved to obtain the constraint. In reality, the systematic uncertainty will dominate the statistical one, and the expected constraint becomes weaker but still as strong as $\Gamma(Z \to \chi \chi) \leq 0.1$\,MeV (or in other words, $\Delta N_\nu \leq 6 \times 10^{-4}$).

\subsection{Electroweak precision measurements}

Here, we discuss the indirect probe of the $Z$-portal WIMP model with the electroweak precision observable (EWPO). As already mentioned in section\,\ref{subsec: setup}, the coupling between the WIMP and the $Z$-boson comes from some heavy mediator particles. The effective interaction after integrating out the mediator can be represented by the dimension-six operator in eq.\,\eqref{eq:effective_interaction}. On the other hand, the mediator particle can also generate other operators in general, which are composed only of SM fields. For instance, the dimension-six Lagrangian ${\cal L}_6$ after integrating the mediator out, could include an effective interaction
\begin{eqnarray}
    \frac{c_H}{\Lambda^2} |H^\dagger D^\mu H|^2.
    \label{eq:OH}
\end{eqnarray}
This interaction, however, contributes to the EWPO and is significantly constrained indeed. This effect is parametrized by the so-called oblique parameters, namely $S,T,U$ parameters\,\cite{Peskin:1990zt, *Peskin:1991sw}. The contribution to the $T$-parameter\,\cite{Ciuchini:2013pca} from the operator in eq.\,\eqref{eq:OH} is
\begin{eqnarray}
    T = -\frac{c_H v^2}{2 \alpha \Lambda^2}.
\end{eqnarray}
The present constraint on the $T$-parameter indicates that $|\Lambda/\sqrt{|c_H|}| \lesssim 6$\,TeV is already excluded at 95\,\% C.L.\,\cite{deBlas:2016ojx,*deBlas:2017wmn}. Therefore the coupling between WIMP and $Z$ boson has the upper-bound $g_{\chi\chi Z} = g_D g_Z v^2/(2 \Lambda^2) \lesssim 6 \times 10^{-4}\,g_D\,c_H^{-1}$. We may expect $c_H = {\cal O}(1)$ and $g_D = {\cal O}(1)$ if both effective interactions \eqref{eq:O6} and \eqref{eq:OH} come from the mediator at tree level. In such a case, the EWPO constraint has a tension with the thermal relic abundance, as $g_{\chi\chi Z} \gg 10^{-3}$ is required for $\Omega_{\rm th} h^2 \leq \Omega_{\rm obs} h^2$. We need small $c_H$ for the relic abundance and the EWPO to be consistent. A way out of this tension is that we may forbid the tree-level effective interaction which contribute to the EWPO. This condition can be satisfied if the mediator is also a $Z_2$ odd particle, as in the case of the supersymmetric bino-Higgsino mixed DM.

There is, however, a loop level contribution to the effective interaction. In fact, though the operator in eq.\,\eqref{eq:effective_interaction} does not contribute to the EWPO, the loop diagram induced from the operator can generate the interaction in eq.\,\eqref{eq:OH}. This loop diagram is quadratically divergent and the loop-induced Wilson coefficient is therefore estimated as follows:
\begin{eqnarray}
    c_H^{\rm loop} \sim \frac{\Lambda_{\rm UV}^2 g_D^2}{16 \pi^2 \Lambda^2},
\end{eqnarray}
where the $\Lambda_{\rm UV}$ is the cut-off of the momentum integration. This loop-induced Wilson coefficient is again constrained by the EWPO. By taking $\Lambda_{\rm UV} \sim \Lambda$, we may get the constraint $|\Lambda/g_D| \lesssim 500$\,GeV. In terms of the coupling between WIMP and the $Z$-boson, the constraint is translated into $g_{\chi \chi Z} \lesssim 0.1$, which is now consistent with the WIMP relic density.

Now let us discuss the future prospect for the indirect probe of the $Z$-portal DM models with the EWPO. The precision measurement at the future lepton colliders can improve the determination of the oblique parameters by an order of magnitude\,\cite{Fan:2014vta}. This prospect indicates that, for the $Z$-portal WIMP model without tree-level EWPO-related operators, the parameter region $g_{\chi \chi Z} \gtrsim 0.01$ can be indirectly tested with the EWPO. This sensitivity can be better than the direct WIMP search at the future lepton colliders, which we will discuss in the next section. Note that although the EWPO-related effective interactions are naturally generated, the precise values of the Wilson coefficients are strongly model-dependent.

We will discuss in the following sections how one can further put a limit on the effective new physics coupling $g_{\chi \chi Z}$ from the WIMP direct production at collider experiments.

\subsection{Constraints from the LEP experiment}

It is also important to discuss the constraint on the fermionic $Z$-portal WIMP DM obtained by the LEP experiment and compare it with the sensitivity of the future lepton colliders. The LEP experiment has also searched for the DM based on the mono-photon signature and obtained the result which is consistent with the prediction of the SM. We therefore use this result to put a constraint on our fermionic $Z$-portal WIMP as given in the following.

We use the data of the 650\,pb$^{-1}$ integrated luminosity collected by the Delphi collaboration during the running of 180--209\,GeV\,\cite{Abdallah:2008aa}. We have generated the Monte-Carlo data of the mono-photon process, $e^- e^+ \to \chi \chi \gamma'{\rm s}$, by using the MadGraph \cite{Alwall:2014hca} and PYTHIA8 \cite{Sjostrand:2007gs} codes. The simplified detector modeling discussed in Ref.\,\cite{Fox:2011fx} was used to take the detector effect into account in the analysis. We compared the Monte-Carlo data (together with the SM background contribution) with the observed data of the single photon energy distribution, and set the upper limit on the $Z$-funnel WIMP signal at 90\,\% C.L. We have only included the statistical uncertainty and neglected the systematic uncertainty.

\subsection{Constraints from LHC searches}

At the LHC, the WIMP search has been performed through processes, where WIMPs are pair produced in association with the SM particles, like photons$(\gamma)$, gluons $(g)$, $W^\pm, Z$, h and quarks. These processes lead to large missing transverse energy in association with mono-photon, jet(s), charged leptons signal at the detector. The theoretical formulation of the DM interaction with the SM sector is based on the assumptions of effective field theory (EFT), where the WIMP interaction is determined by the Lorentz structure of the effective interaction parameterized by the WIMP mass and the cut-off scale $\Lambda $. The cut off scale is the relevant scale of the process obtained after integrating out the heavy mediator particle. In an alternative approach, the so-called simplified scenario is assumed where the mass of the  mediator particle is well within the kinematic reach of the LHC and the WIMP interaction with the ordinary matter is determined by the Lorentz structure, WIMP mass, mediator mass along with the couplings of the mediator with the WIMP $(g_\chi)$ and the standard model particles $(g_{q})$. At the LHC, while looking for the signature of the WIMPs, it is generally assumed that the WIMP mass is relatively small compared to the mass of the mediator. The exclusion limit from null observation is usually given in terms of the mediator particle mass and the WIMP mass assuming particular values of  $g_\chi$ and $g_q$ in a simplified model. This exclusion limit can be easily translated in the effective field theory scenario in terms of the WIMP mass and the cut-off scale. Latest constraints from Run-II of the LHC experiment at 13\,TeV are given by the ATLAS collaboration from mono-photon search\,\cite{Aaboud:2017dor} and mono-jet searches\,\cite{Aaboud:2017phn}. A detailed likelihood analysis on the $Z$-portal DM model with a singlet Majorana DM including exclusion limits from LHC mono-photon and mono-jet searches has shown that only the new physics coupling greater than ${\cal O}$(1) is excluded at 95\,\% C.L. for the WIMP mass at the $Z$-funnel region\,\cite{Balazs:2017ple}. However, this limit is less stringent than the previous LEP constraint and thus shall not be further discussed in our analysis.

\section{Results}
\label{sec: result}

We will discuss the capability of the future lepton colliders in searching the fermionic $Z$-portal WIMP based on our result obtained in the previous section. We first discuss the beam bremsstrahlung and initial state radiation effects as well as the detector effect to make our analysis realistic to some extent. Then, we present the region which could be covered by the future lepton colliders through the mono-photon search and the precise measurement of the invisible $Z$ boson decay width. These results are compared with constraints obtained by the LEP experiment and the direct DM detection at underground experiments.

\subsection{Various effects on lepton collider experiments}
\label{secsec: the effects}

\subsubsection{Beam bremsstrahlung effect}

When the electron and positron beams collide with each other, the energy distribution of each beam bunch is not monochromatic at the initial beam energy but described by the function having a long tail at the low energy region. This is because, when the electron and positron beams are closer, the beams come under the influence of the electromagnetic fields of each other and lose their energies through the bremsstrahlung process. We involve this beam bremsstrahlung effect using the formula developed in Refs.\,\cite{Chen:1988pn, Peskin:1999pk, Datta:2005gm}. With $x \equiv E/E_{in}$, the formula gives us the energy distribution function of the electrons and positrons as
\begin{eqnarray}
	\psi_e(x) = e^{-N_\gamma} \left[ \delta(x - 1) + \frac{ e^{-\eta_x} } {x (1 - x) }\,h(N_\gamma \eta_x^{1/3}) \right] 
	,
\end{eqnarray}
where the distribution function is normalized to be $\int^1_0 dx\,\psi_e(x) = 1$. The two parameters in the above formula are given by $N_\gamma = \sqrt{3} \sigma_z \nu _{cl} (1 + \Upsilon^{2/3})$ and $\eta_x = (1/x - 1)\,\kappa$, respectively, with $\nu_{cl} = 5 \alpha^2 \Upsilon/(2 \sqrt{3}\,r_e \gamma_0)$, $\Upsilon = 5 r_e \gamma_0 N/\{6 \alpha \sigma_z (\sigma_x +\sigma_y)\}$ and $\kappa = 2/(3\Upsilon)$. Here, $\alpha$ is the fine structure constant, $r_e \simeq 2.82 \times 10^{-15}$\,m is the classical electron radius, $\gamma_0 = E_{in}/(m_e c^2)$ with $m_e$ being the electron mass and $N \simeq 2 \times 10^{10}$ is the total number of electrons/positrons in a bunch. Beams sizes are fixed to be $\sigma_x = 729$\,nm, $\sigma_y = 7.7$\,nm and $\sigma_z = 0.3$\,mm, respectively, referring to the values in the technical design report of the ILC experiment\,\cite{Behnke:2013xla}. See also Ref.\,\cite{Datta:2005gm} for the concrete form of the function ``$h$''. For the case of CEPC experiment, we can ignore this effect, because the shape of the beam is enough broad.

\subsubsection{Initial state radiation effect}

Initial state radiation (ISR) is the effect that the incident electron and positron emit soft photons just before the collision so that the beam energy (the collision energy, as a result) effectively diminishes. Among various methods to take the ISR effect into account, we adopt the one that was used for the mono-photon search at the LEP experiment\,\cite{Kuraev:1985hb}. Here, the ISR effect is involved through the function $F(x)$ with $x$ being $x = E_{\rm out}/E_{\rm in}$, and the function $F(x)$ gives the energy ($E_{\rm out}$) distribution of the electron or positron (that originally has the energy of $E_{\rm in}$) after it experienced the ISR. Its explicit form is given by the formula:
\begin{eqnarray}
	F(x) = e^{\beta (3/4 - \gamma_E)} \beta (1 - x)^{\beta - 1}
	\frac{ 4(1 + x^2) - \beta \left[ (1 + 3 x^2) \ln x + 2 (1 - x)^2 \right] }{8 \Gamma(1 + \beta)},
\end{eqnarray}
where $\gamma_E \simeq 0.577$ is the Euler constant, $\beta = \alpha\,[2\log(E_{\rm in}/m_e) - 1]/\pi$ and $\Gamma(x)$ is the gamma function. We convolute the ISR effect with the beam bremsstrahlung effect to obtain the realistic collision energy used to produce a pair of WIMPs (with an energetic photon; $E_\gamma>10$ GeV and $|\cos{\theta_\gamma}| < 0.98$).

\subsubsection{Detector effect}

An energetic photon produced by the signal or background process is detected at the electromagnetic calorimeter, in which the photon causes an electromagnetic shower, creating a cascade of electron-positron pairs and bremsstrahlung photons. The electrons and positrons lose energy through ionization and are eventually stopped. The calorimeter measures this energy loss, which enables us to measure the energy of the original photon. Since the measurement owes to the stochastic process, it associates with an uncertainty caused by the number fluctuation in the development of cascade showers. This effect emerges as the stochastic term in the resolution of the calorimeter, and it is proportional to the square root of the photon energy due to the property of the fluctuation. In addition, there is another term in the resolution called the constant term, which comes from the calibration.

The resolution of the electromagnetic calorimeter is therefore given by the form:
\begin{eqnarray}
	\frac{ \sigma( E_\gamma ) } { E_\gamma } = \frac{ a }{ \sqrt{E_\gamma} } + b,
\end{eqnarray}
where $E_\gamma$ is the photon energy coming into the calorimeter, $\sigma({E_\gamma})$ is the resolution of calorimeter which depends on $E_\gamma$, `$a$' is the stochastic term and `$b$' is the constant term. At the ILC experiment, the values of the constants are expected to be $a = 0.166$ and $b = 0.011$ (in the unit of GeV)\,\cite{Adloff:2008aa}. We adopt these values and take the detector effect into account in our analysis by smearing the energy of the photon at the final state.

\subsubsection{Cross sections including all the effects}

\begin{figure}[t!]
	\centering
	\includegraphics[height=2.5in, angle=0]{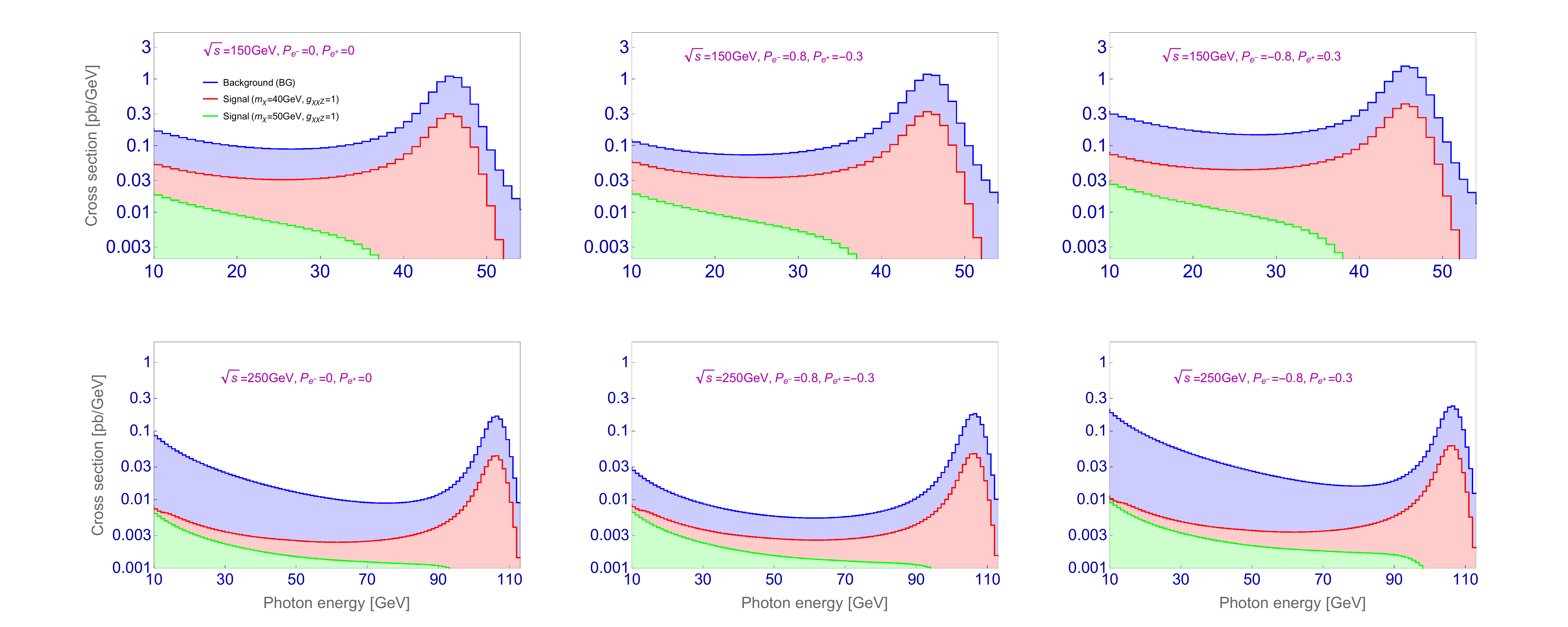}
	\caption{\small \sl Differential cross sections for the signal with $m_\chi = 40$\,GeV and the signal with $m_\chi = 50$\,GeV and the irreducible background including all the beam bremsstrahlung, initial state radiation and detector effects. The other parameters are fixed to be the same as those in Fig.\,\ref{fig: cross section}.}
	\label{fig: cross section II}
\end{figure}

Differential cross sections, which are the same as those in Fig.\,\ref{fig: cross section} but involves all the beam bremsstrahlung, initial state radiation and detector effects, are shown in Fig,\,\ref{fig: cross section II}. It is seen from the comparison between the two figures (Figs.\,\ref{fig: cross section} and \ref{fig: cross section II}) that the peak structure at around $E_\gamma = (s - m_Z^2)/(2 \sqrt{s})$ is smeared due to the effects, while the cross section below the peak is enhanced. This is because the collision energy can be below $\sqrt{s}$ due to the beam bremsstrahlung and initial state radiation effects, and hence the on-shell $Z$ boson production becomes possible by emitting a photon with an energy less than $(s - m_Z^2)/(2 \sqrt{s})$ and it is well known as the return of the $Z$-peak. 

\subsection{Capability of the future lepton colliders}

\begin{figure}[t!]
	\centering
	\includegraphics[height=2.7in, angle=0]{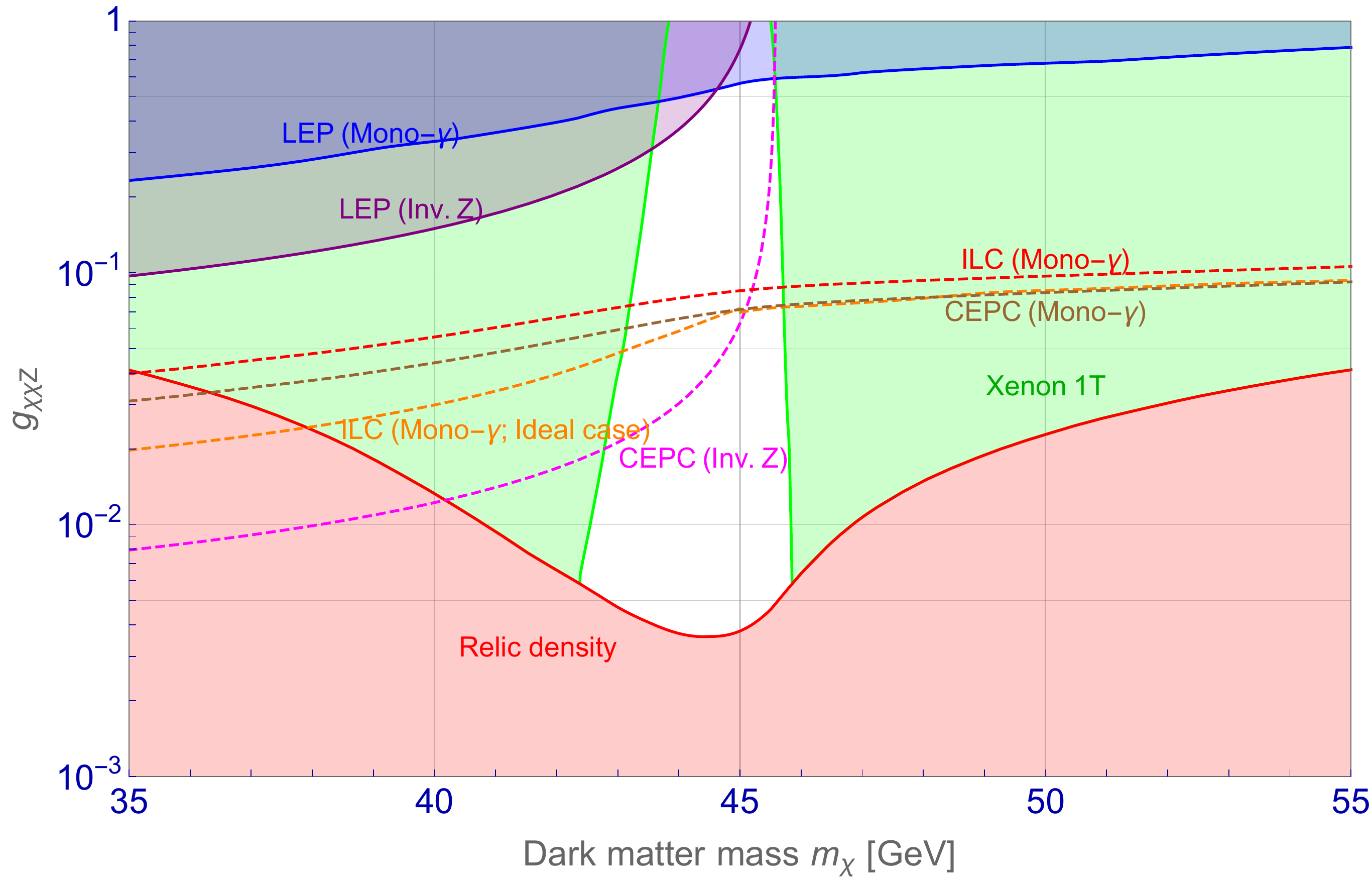}
	\caption{\small \sl The shaded regions show the various constraints on the $Z$-portal WIMP model on the $(m_\chi, g_{\chi \chi Z})$-plane. These are the relic density (red), the direct detection (green) and the LEP Mono-photon and Invisible Z-width constraints (blue and violet respectively). The sensitivities of the future lepton colliders are shown as broken lines. The labelling is self-explanatory. See the text for more details.}
	\label{fig: Stat_only}
\end{figure}

Our results are summarized on the $(m_\chi, g_{\chi \chi Z})$-plane in Fig.\,\ref{fig: Stat_only}. The present constraints at 90\,\% C.L. from the LEP experiments (invisible $Z$ decay width search and mono-photon search), the relic abundance observation and the direct DM detection experiments of the WIMP discussed in section\,\ref{subsec: relic abundance} are shown as shaded regions. On the other hand, the sensitivity of the future lepton colliders (ILC and CEPC), which is defined as the future expected constraint at 90\,\% C.L. if no WIMP signal is detected there, are shown as several broken lines: One named "CEPC (Inv. $Z$)" is from the invisible decay width search of the $Z$ boson at the CEPC experiment (240\,GeV \& 5\,ab$^{-1}$; no polarization assumed), while the other named "CEPC (Mono-$\gamma$)" is from the mono-photon search at the same experiment. The line named "ILC (Mono-$\gamma$)" is from the mono-photon search at the ILC experiment (250\,GeV \& 2\,ab$^{-1}$; 1\,ab$^{-1}$ for each polarization), and that named "ILC (Mono-$\gamma$; Ideal case)" is from the same mono-photon search but assuming the optimized center-of-mass energy and polarization at each DM mass as discussed in section\,\ref{subsubsec: optimization}.\footnote{We note that, when the WIMP mass is less than 45\,GeV, the optimized center-of-mass energy is shifted to a higher value than that shown in Fig.\,\ref{fig: optimization} due to beam and detector effects discussed in section\,\ref{secsec: the effects}. We have computed the optimized energy again including the effects and used it to depict the sensitivity line.}\footnote{The points on this sensitivity line cannot be simultaneously achieved, because each point corresponds to a data set with a different center-of-mass energy. Thus, it should be understood as the sensitivity that the ILC experiment ideally has when we have a hint of the DM mass at some other experiments (e.g.  direct DM detection). It is also worth pointing out that, the instantaneous luminosity typically becomes lower for lower center-of-mass energies at linear colliders, though the same integrated luminosity is assumed along the sensitivity line. Thus, the time needed for the data taking is different along the line. It should be also noted that circular colliders have greater luminosity as compared to linear ones at lower center-of-mass energies.} We only consider statistical errors and use the likelihood analysis discussed in section\,\ref{subsubsec: optimization} to depict the sensitivity lines.

It can be seen in the figure that the future lepton colliders can play an important role to investigate an uncharted parameter region. It is also worth pointing out that the entire uncharted parameter region of the $Z$-portal DM will be covered by future direct DM detection experiments. Hence, once the DM signal is detected in a certain parameter region, the scientific significance of the future lepton colliders will be significantly increased.

The sensitivity of the future lepton colliders in Fig.\,\ref{fig: Stat_only} is estimated including only statistical uncertainties. On the other hand, for more realistic estimate, we have to take into account systematic experimental uncertainties $(\delta B)$. Since it is difficult to evaluate the systematic uncertainties rigorously before the experiments start, so we discuss its effect on the sensitivity of the future lepton colliders assuming $\delta B = $1\,\% and 0.1\,\%. To include the systematic uncertainty, we use an appropriate single energy bin to evaluate the likelihood, unlike the small binned analysis in section\,\ref{subsubsec: optimization}. This is because the systematic uncertainties may have a correlation between the bins. The target energy bin is selected so that the photon energy is in between 10\,GeV to $X$\,GeV with $X$ being determined to maximize the significance at each DM mass. Then, the likelihood is defined by the following equation:
\begin{eqnarray}
	\Delta \chi^2 = \Max
	\left( \left. \frac{ \{ N(X) - N^{\rm BG}(X) \}^2 } {\delta N^{\rm BG}(X)^2 + N^{\rm BG}(X)} \right| 10\,{\rm GeV} \leq E_\gamma \leq X\,{\rm GeV} \right).
	\label{eq: systematic uncertainty delta chi2}
\end{eqnarray}
Here, $N(X)$ is the expected number of signal plus background events in between 10\,GeV to $X$\,GeV, while $\delta N^{\rm BG}(X)$ concerns the systematic uncertainty, which will be estimated as a product of a given uncertainty (0.1\,\% or 1\,\%) and the number of background events.

\begin{figure}[t!]
	\centering
	\includegraphics[height=2.7in, angle=0]{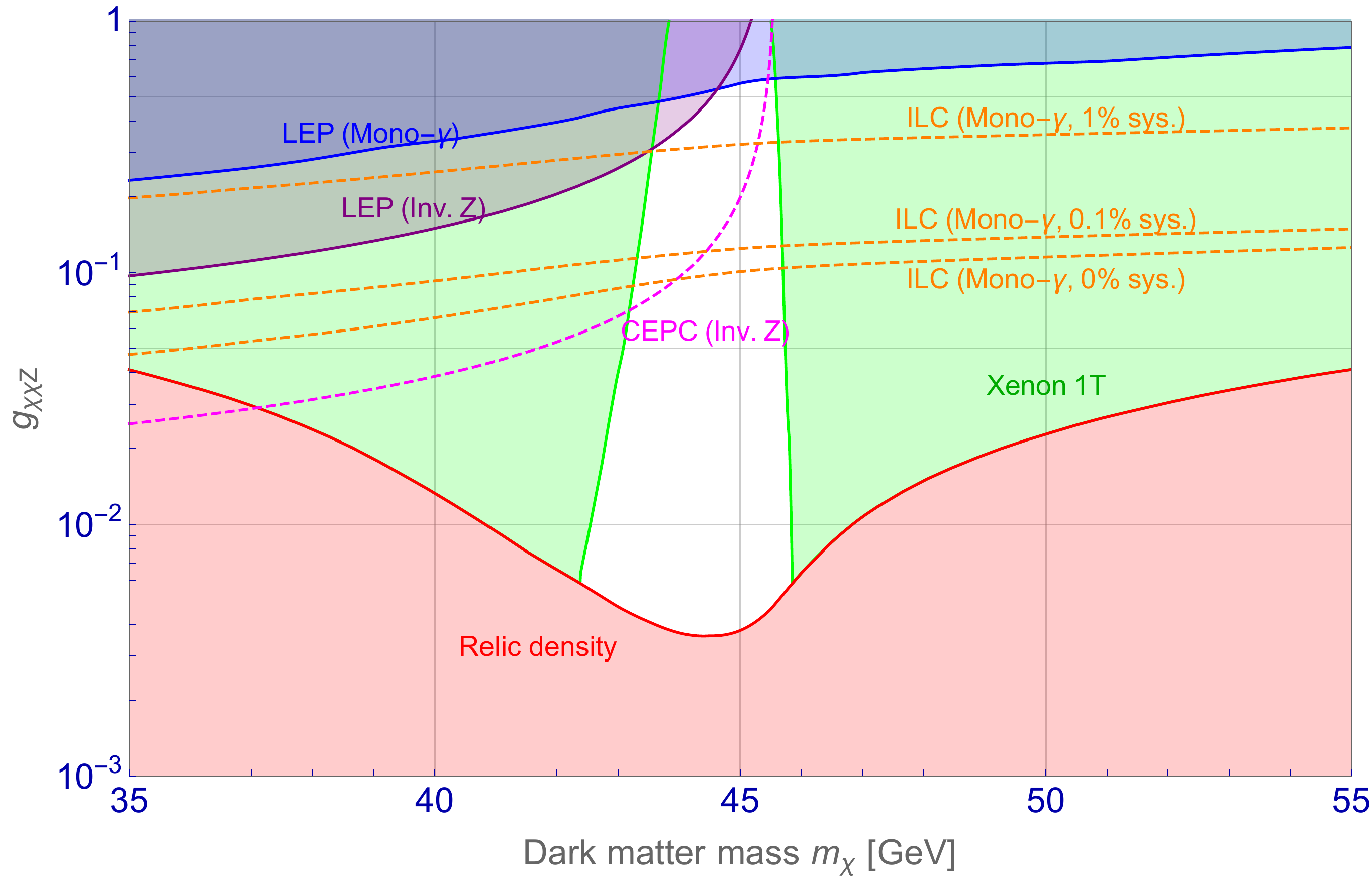}
	\caption{\small \sl The color definintion is the same as Fig.\ref{fig: Stat_only} except the future collider sensitivity includes systematic uncertainty. See the text for more details.}
	\label{fig: systematic uncertainty}
\end{figure}

The result is shown in Fig.\,\ref{fig: systematic uncertainty}, where the sensitivity of the mono-photon search at the ILC experiment (250\,GeV \& 2\,ab$^{-1}$; 1\,ab$^{-1}$ for each polarization) with the systematic uncertainties of 0\,\%, 0.1\,\% and 1\,\% is shown.\footnote{The reason why the sensitivity line with 0\,\% systematic uncertainty is different from the one in Fig.\,\ref{fig: Stat_only} is we use different likelihoods. The likelihood in eq.\,(\ref{eq: systematic uncertainty delta chi2}) with $\delta N^{\rm BG}(X)=0$ is less sensitive than that of Fig.\,\ref{fig: Stat_only}.} The sensitivity of the invisible decay width search of the $Z$ boson at the CEPC experiment is also shown with the systematic uncertainty addressed in section\,\ref{subsec: Inv Z}. It is seen that the systematic uncertainty of ${\cal O}$(0.1)\,\% is not very different from the one without the uncertainty, while the uncertainty of ${\cal O}(1)$\,\% makes the sensitivity significantly worse. Handling the uncertainty at ${\cal O}(0.1)$\,\% level is thus mandatory to make the future lepton collider sensitive enough to search for the $Z$-portal DM.

\section{Conclusions}

We have presented the expected sensitivity of the $Z$-portal WIMP at the future lepton colliders. We have adopted the effective operator method where the interaction between the singlet Majorana WIMP $(\chi)$ and the $Z$ boson is mediated via the dimension-six operator $(\bar{\chi} \gamma_\mu \gamma_5 \chi) (H^\dagger i D^{\mu}H)/2+h.c$. The final result of our analysis thus is parametrized by only two parameters, the WIMP mass $(m_\chi)$ and the effective WIMP-$Z$ coupling $(g_{\chi \chi Z})$.

Through this work, we discussed the possibility of probing the $Z$-funnel WIMP mass region (35-55 GeV) using the mono-photon plus missing energy signal at the future lepton colliders (ILC \& CEPC). We have done a comprehensive signal-background analysis of the mono-photon searches considering various collider features, such as beam polarization, beam breamsstrahlung, initial-state-radiation and detector effects. While doing this analysis we have taken into account other important constraints on the parameters of this scenario coming from the mono-photon searches, $Z$-invisible width and EW precision measurement obtained from the LEP data. 

For the mono-photon signal, the dominant irreducible background comes from neutrino pair production with additional photons due to initial state radiations. The primary distinction between the signal and the background can be made by observing the energy and angular distribution of the single photon. We have done a $\Delta\chi^2$ analysis to calculate the signal sensitivity over the background from the differential photon energy distribution. This has been done for three different choices of beam polarization. The optimized centre of mass of energy for the beams have been determined by the maximal $\Delta\chi^2$ value for each polarization. As expected, we have found that the right-handed beam polarization works better for WIMP masses greater than $m_Z/2$ while the left-handed beam polarization yield better sensitivity for $m_\chi < m_Z/2$. Further we showed that the inclusion of ISR effect, beam bremsstrahlung and the detector effect on the photon energy smears the peak structure of the energy distribution. Therefore, for the likelihood analysis, the numbers of the signal and background events for each particular bin of the photon energy differs and hence, these effects contribute to the final signal significance. We have done a complete analysis including these effects in calculating the future discovery reach of the lepton colliders.

 We studied the prospects of the $Z$-funnel WIMP detection at the ILC and CEPC detectors by using the optimized beam energy and polarization obtained from the $\Delta\chi^2$ analysis. The 90\,\% C.L. limit is shown on the $(m_\chi, g_{\chi \chi Z})$-plane for ILC at 250\,GeV and 2\,ab$^{-1}$ luminosity and for CEPC at 240\,GeV and 5\,ab$^{-1}$ luminosity. Furthermore, this limit is combined with the ones obtained from the $Z$-invisible width measurement, direct detection and relic abundance of the WIMP. Additionally, we have done a realistic estimation including the systematic uncertainties for the ILC beam. To do so, we re-estimate the $\Delta\chi^2$ with 0.1\,\% and 1\,\% systematic uncertainties. The collective 90\,\% C.L. bound for all the cases for a 250\,GeV ILC beam with 0\,\%, 0.1\,\% and 1\,\% systematic uncertainties has also been presented.  

 It is important to mention that the direct DM search experiments such as XENON1T and PICO-60 have put severe limits on $Z$-portal WIMPs. In fact, if the WIMP $\chi$ is a dominant component of the DM and we adopt the assumption of the standard local DM density $\rho_{\rm DM} = 0.3~{\rm GeV/cm^3}$, the current constraint from the XENON1T has already excluded the reach of the future lepton colliders. As discussed in Sec. \ref{subsec:DD}, however, the direct DM search constraints critically depend on these assumptions. For instance, if the coupling $g_{\chi\chi Z}$ is larger and the $\chi$ is only a subdominant component of the DM, the present direct DM search cannot cover the mass range $m_{\chi} \in [42, 46]$ GeV. There is a potentially large astrophysical uncertainty in the local DM density, which also significantly affects the reach of the DM direct detection experiments. The future lepton colliders can probe the WIMP DM without such uncertainties. 
 
 We also studied the future prospect of direct detection, and found that it is possible to discover $Z$-portal WIMP even if we assume very conservative cosmological set up. If future direct detection experiments discover a WIMP, the role of the future lepton colliders is particularly important, as it provide unique opportunity to identify the character of the WIMP. In conclusion, this study reveals the prospect of WIMP detection in future lepton colliders, which is the most conservative test and confirmation of the  $Z$-portal WIMP model.

\appendix
\section{Appendix}
\label{app: direct detection}

Following the discussion in Sec.\,\ref{subsec:DD}, here we present a detailed analysis of the spin-dependent WIMP-nucleon scattering cross section measurement at the direct detection experiment for our $Z$-portal WIMP scenario.
\begin{figure}[t!]
	\centering
	\includegraphics[height=2.7in, angle=0]{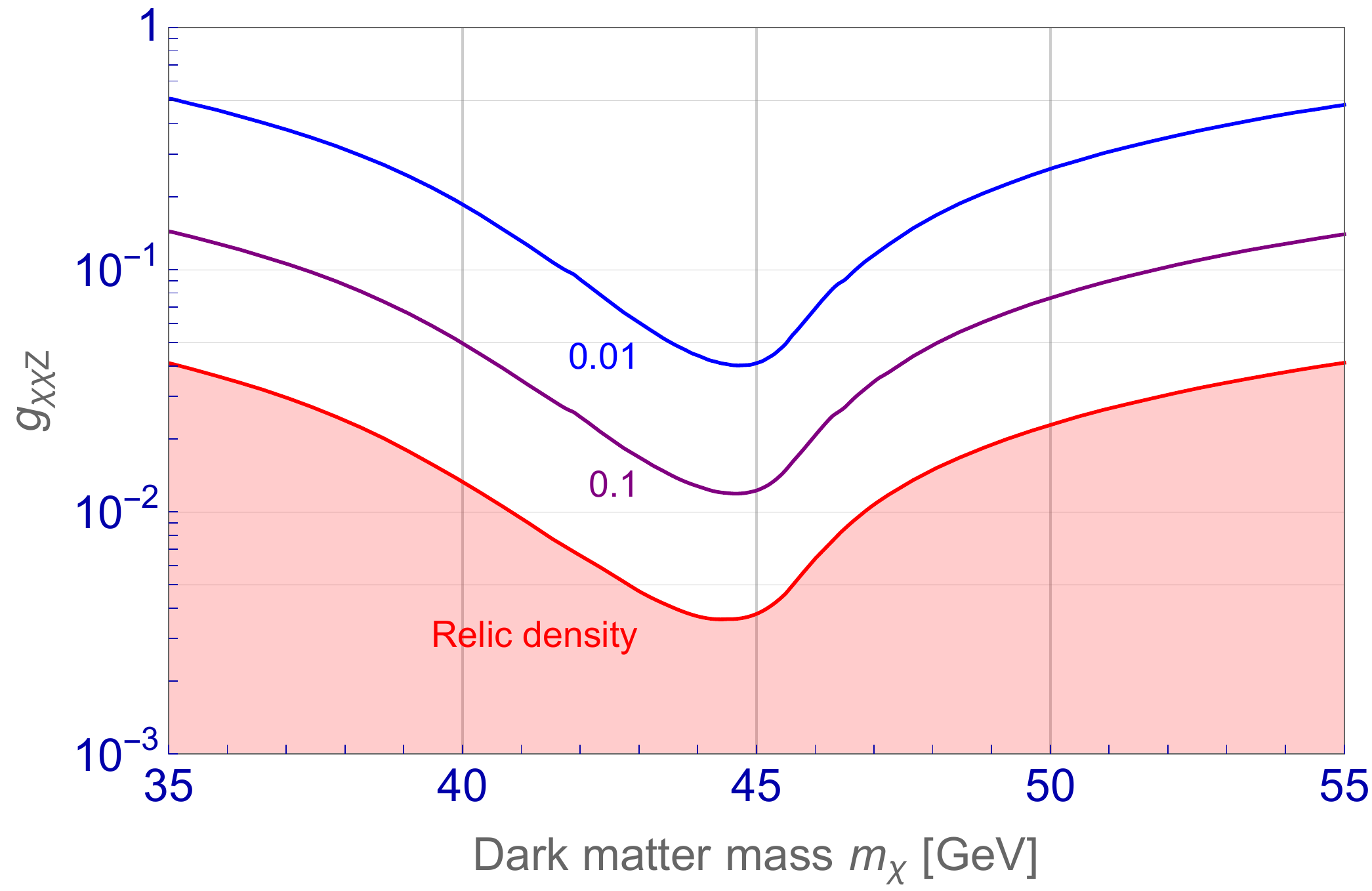}
  \caption{\small \sl The 2$\sigma$ allowed region from the relic density constraint (red solid line). The contours on the white region show the contribution of this WIMP to the total DM abundance in that parameter space.}
  \label{fig:abundance}
\end{figure}
As already mentioned, the scattering cross section of WIMP off nuclei is inherently related to the WIMP relic abundance because the same interaction governs both the processes. In this study, we mainly concentrate on the scenario where our $Z$-portal WIMP is an under-abundant DM candidate and can only partly contribute to the total DM density. In Fig.\,\ref{fig:abundance}, we show the contours of fractions of relic density contributed by the WIMP to the total DM density in the model parameter space where it is under-abundant (the white region above the red-shaded region). 

Now, the scattering rate of the WIMP also includes the DM halo density. Therefore, if the proposed WIMP candidate constitutes only a part of the total DM density then the scattering rate of the WIMP is scaled by the halo fraction of the WIMP which will simply be the fraction of the WIMP abundance to the total DM density, as given by eq.\,(\ref{eq:effDD}). Therefore, following eq.\,(\ref{eq:effDD}), the correct WIMP-nucleon scattering cross section in that particular parameter space should be scaled by this fraction of abundance. 
\begin{figure}[htbp]
	\centering
	\includegraphics[width=75mm,height=60mm, angle=0]{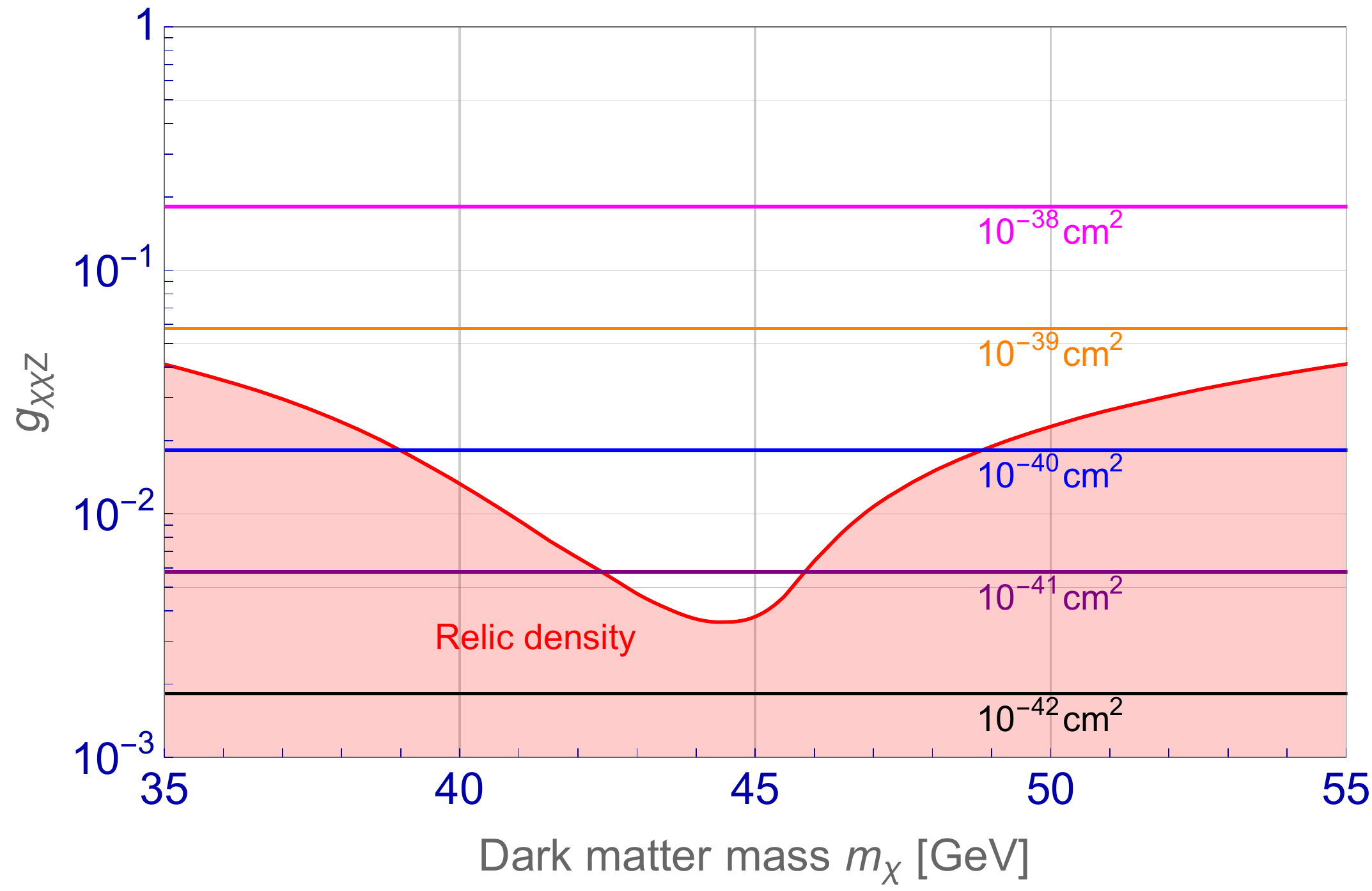} 
	\includegraphics[width=75mm,height=60mm, angle=0]{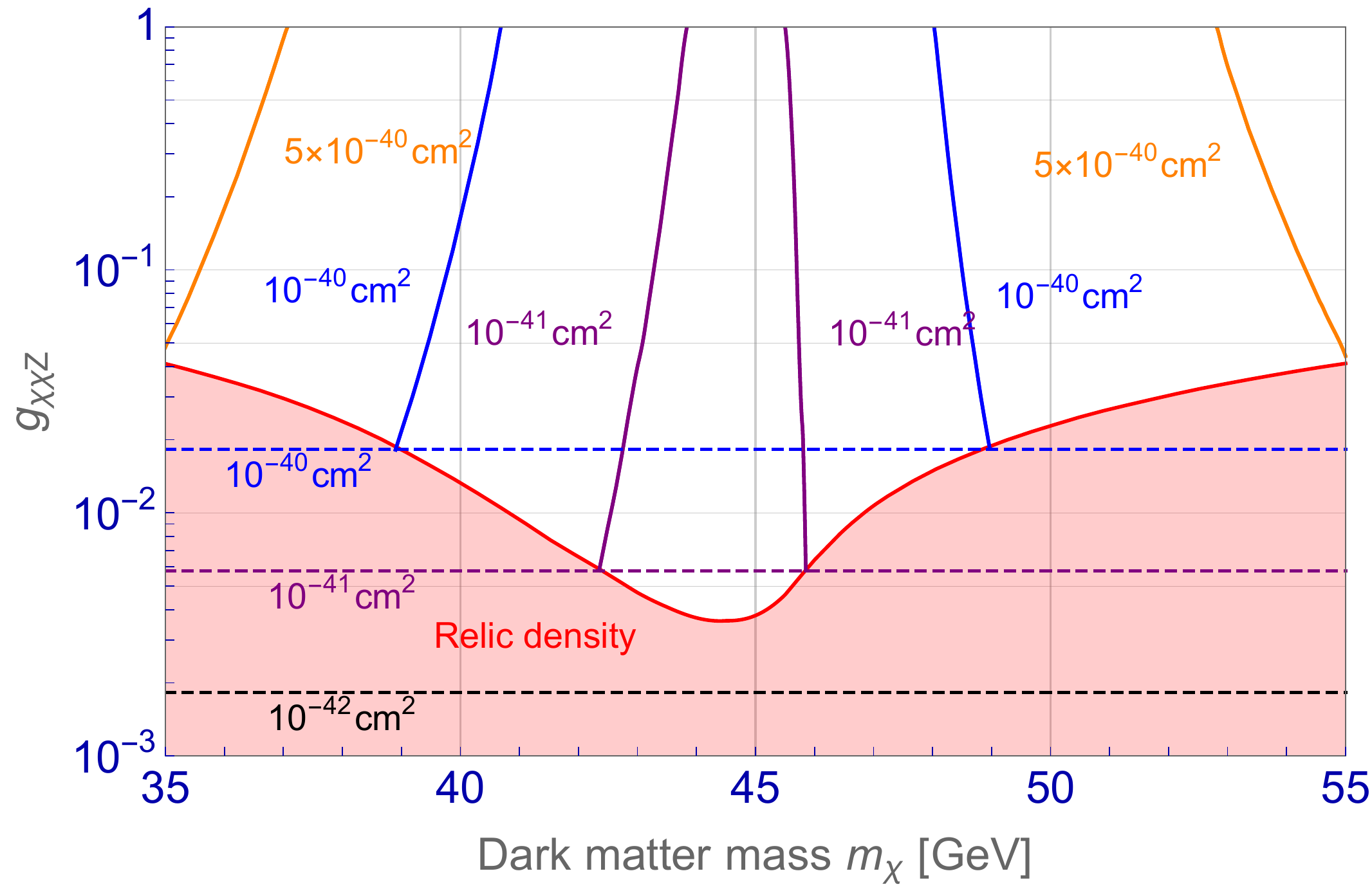}
	\caption{\small \sl  Contours show un-scaled (Left) and scaled (Right) WIMP-neutron scattering cross section over the whole parameter space. Relic density constraint follow the same color definition as in Fig,\ref{fig: relic abundance}.}
	\label{fig: direct detection}
\end{figure}

In the left panel of Fig.\,\ref{fig: direct detection}, we show the un-scaled scattering cross section for our WIMP candidate in the ($m_{\chi}-g_{\chi \chi Z}$) plane. The un-scaled scattering cross section has no dependence on the mass $m_{\chi}$ of the WIMP as also mentioned in eq.\,(\ref{eq:unscaledDD}). We show four contours of cross section value and the one with value $10^{-41}$\,cm$^2$ corresponds to the latest WIMP-neutron cross section limit by the XENON1T\,\cite{Aprile:2019dbj} experiment, the area above which is excluded. It is evident that the current XENON1T bound still allows a small parameter space near the $Z$-funnel for a thermal WIMP candidate where the red solid line in the figure refer to the $2\sigma$ allowed region by the PLANCK relic density constraint. The contour with value $10^{-42}$\,cm$^2$ corresponds to the
conservative detectability of future LZ\,\cite{Mount:2017qzi} direct detection experiment, therefore the entire parameter region can be covered by this experiment.    

In the right panel of Fig.\,\ref{fig: direct detection}, the contours of the scaled scattering cross sections are plotted as solid line. The cross section is scaled by the fraction of the WIMP abundance following Fig.\,\ref{fig:abundance} in the specific parameter space. In analogy to our previous statement in Sec.\,\ref{subsec:DD}, one can evidently see that the scaled scattering cross section is approximately independent of the effective coupling $g_{\chi \chi Z}$. The un-scaled cross section contours has also been overlapped as dashed lines. Following the current bound from XENON1T, the combined region by both the scaled and un-scaled contours of value $10^{-41}$\,cm$^2$ thus describes the current reach of DM direct detection experiment in this particular WIMP scenario. This also explains the green shaded region excluded by XENON1T in Figs.\,\ref{fig: relic abundance}, \ref{fig: Stat_only} and \ref{fig: systematic uncertainty}.

\vspace{0.3cm}
\noindent
{\bf Acknowledgments}
\vspace{0.1cm}\\
\noindent
This work is supported by Grant-in-Aid for Scientific Research from the Ministry of Education, Culture, Sports, Science, and Technology (MEXT), Japan, No.~16H02176, 19H05810 (S.M.), 17H02878 (S.M. and S.S.), 18K13535, 19H04609 (S.S.), and by World Premier International Research Center Initiative (WPI), MEXT, Japan. DKG acknowledges the hospitality of the Kavli IPMU, University of Tokyo where this work was initiated and the Theoretical Physics Department, CERN, Switzerland where part of this work was done. 

\bibliographystyle{aps}
\bibliography{refs}

\begin{thebibliography}{56}%
\makeatletter
\providecommand \@ifxundefined [1]{%
 \@ifx{#1\undefined}
}%
\providecommand \@ifnum [1]{%
 \ifnum #1\expandafter \@firstoftwo
 \else \expandafter \@secondoftwo
 \fi
}%
\providecommand \@ifx [1]{%
 \ifx #1\expandafter \@firstoftwo
 \else \expandafter \@secondoftwo
 \fi
}%
\providecommand \natexlab [1]{#1}%
\providecommand \enquote  [1]{``#1''}%
\providecommand \bibnamefont  [1]{#1}%
\providecommand \bibfnamefont [1]{#1}%
\providecommand \citenamefont [1]{#1}%
\providecommand \href@noop [0]{\@secondoftwo}%
\providecommand \href [0]{\begingroup \@sanitize@url \@href}%
\providecommand \@href[1]{\@@startlink{#1}\@@href}%
\providecommand \@@href[1]{\endgroup#1\@@endlink}%
\providecommand \@sanitize@url [0]{\catcode `\\12\catcode `\$12\catcode
  `\&12\catcode `\#12\catcode `\^12\catcode `\_12\catcode `\%12\relax}%
\providecommand \@@startlink[1]{}%
\providecommand \@@endlink[0]{}%
\providecommand \url  [0]{\begingroup\@sanitize@url \@url }%
\providecommand \@url [1]{\endgroup\@href {#1}{\urlprefix }}%
\providecommand \urlprefix  [0]{URL }%
\providecommand \Eprint [0]{\href }%
\providecommand \doibase [0]{http://dx.doi.org/}%
\providecommand \selectlanguage [0]{\@gobble}%
\providecommand \bibinfo  [0]{\@secondoftwo}%
\providecommand \bibfield  [0]{\@secondoftwo}%
\providecommand \translation [1]{[#1]}%
\providecommand \BibitemOpen [0]{}%
\providecommand \bibitemStop [0]{}%
\providecommand \bibitemNoStop [0]{.\EOS\space}%
\providecommand \EOS [0]{\spacefactor3000\relax}%
\providecommand \BibitemShut  [1]{\csname bibitem#1\endcsname}%
\let\auto@bib@innerbib\@empty
\bibitem [{\citenamefont {Aad}\ \emph {et~al.}(2012)\citenamefont {Aad} \emph
  {et~al.}}]{Aad:2012tfa}%
  \BibitemOpen
  \bibfield  {author} {\bibinfo {author} {\bibfnamefont {G.}~\bibnamefont
  {Aad}} \emph {et~al.} (\bibinfo {collaboration} {ATLAS}),\ }\href {\doibase
  10.1016/j.physletb.2012.08.020} {\bibfield  {journal} {\bibinfo  {journal}
  {Phys. Lett.}\ }\textbf {\bibinfo {volume} {B716}},\ \bibinfo {pages} {1}
  (\bibinfo {year} {2012})},\ \Eprint
  {http://arxiv.org/abs/1207.7214}{arXiv:1207.7214 [hep-ex]}\BibitemShut
  {NoStop}%
\bibitem [{\citenamefont {Chatrchyan}\ \emph {et~al.}(2012)\citenamefont
  {Chatrchyan} \emph {et~al.}}]{Chatrchyan:2012ufa}%
  \BibitemOpen
  \bibfield  {author} {\bibinfo {author} {\bibfnamefont {S.}~\bibnamefont
  {Chatrchyan}} \emph {et~al.} (\bibinfo {collaboration} {CMS}),\ }\href
  {\doibase 10.1016/j.physletb.2012.08.021} {\bibfield  {journal} {\bibinfo
  {journal} {Phys. Lett.}\ }\textbf {\bibinfo {volume} {B716}},\ \bibinfo
  {pages} {30} (\bibinfo {year} {2012})},\ \Eprint
  {http://arxiv.org/abs/1207.7235}{arXiv:1207.7235 [hep-ex]}\BibitemShut
  {NoStop}%
\bibitem [{\citenamefont {Behnke}\ \emph {et~al.}(2013)\citenamefont {Behnke},
  \citenamefont {Brau}, \citenamefont {Foster}, \citenamefont {Fuster},
  \citenamefont {Harrison}, \citenamefont {Paterson}, \citenamefont {Peskin},
  \citenamefont {Stanitzki}, \citenamefont {Walker},\ and\ \citenamefont
  {Yamamoto}}]{Behnke:2013xla}%
  \BibitemOpen
  \bibfield  {author} {\bibinfo {author} {\bibfnamefont {T.}~\bibnamefont
  {Behnke}}, \bibinfo {author} {\bibfnamefont {J.~E.}\ \bibnamefont {Brau}},
  \bibinfo {author} {\bibfnamefont {B.}~\bibnamefont {Foster}}, \bibinfo
  {author} {\bibfnamefont {J.}~\bibnamefont {Fuster}}, \bibinfo {author}
  {\bibfnamefont {M.}~\bibnamefont {Harrison}}, \bibinfo {author}
  {\bibfnamefont {J.~M.}\ \bibnamefont {Paterson}}, \bibinfo {author}
  {\bibfnamefont {M.}~\bibnamefont {Peskin}}, \bibinfo {author} {\bibfnamefont
  {M.}~\bibnamefont {Stanitzki}}, \bibinfo {author} {\bibfnamefont
  {N.}~\bibnamefont {Walker}}, \ and\ \bibinfo {author} {\bibfnamefont
  {H.}~\bibnamefont {Yamamoto}},\ }\href@noop {} {\  (\bibinfo {year}
  {2013})},\ \Eprint {http://arxiv.org/abs/1306.6327}{arXiv:1306.6327
  [physics.acc-ph]}\BibitemShut {NoStop}%
\bibitem [{\citenamefont {Boehm}\ \emph
  {et~al.}(2004{\natexlab{a}})\citenamefont {Boehm}, \citenamefont {Ensslin},\
  and\ \citenamefont {Silk}}]{Boehm:2002yz}%
  \BibitemOpen
  \bibfield  {author} {\bibinfo {author} {\bibfnamefont {C.}~\bibnamefont
  {Boehm}}, \bibinfo {author} {\bibfnamefont {T.~A.}\ \bibnamefont {Ensslin}},
  \ and\ \bibinfo {author} {\bibfnamefont {J.}~\bibnamefont {Silk}},\ }\href
  {\doibase 10.1088/0954-3899/30/3/004} {\bibfield  {journal} {\bibinfo
  {journal} {J. Phys.}\ }\textbf {\bibinfo {volume} {G30}},\ \bibinfo {pages}
  {279} (\bibinfo {year} {2004}{\natexlab{a}})},\ \Eprint
  {http://arxiv.org/abs/astro-ph/0208458}{arXiv:astro-ph/0208458
  [astro-ph]}\BibitemShut {NoStop}%
\bibitem [{\citenamefont {Boehm}\ \emph
  {et~al.}(2004{\natexlab{b}})\citenamefont {Boehm}, \citenamefont {Hooper},
  \citenamefont {Silk}, \citenamefont {Casse},\ and\ \citenamefont
  {Paul}}]{Boehm:2003bt}%
  \BibitemOpen
  \bibfield  {author} {\bibinfo {author} {\bibfnamefont {C.}~\bibnamefont
  {Boehm}}, \bibinfo {author} {\bibfnamefont {D.}~\bibnamefont {Hooper}},
  \bibinfo {author} {\bibfnamefont {J.}~\bibnamefont {Silk}}, \bibinfo {author}
  {\bibfnamefont {M.}~\bibnamefont {Casse}}, \ and\ \bibinfo {author}
  {\bibfnamefont {J.}~\bibnamefont {Paul}},\ }\href {\doibase
  10.1103/PhysRevLett.92.101301} {\bibfield  {journal} {\bibinfo  {journal}
  {Phys. Rev. Lett.}\ }\textbf {\bibinfo {volume} {92}},\ \bibinfo {pages}
  {101301} (\bibinfo {year} {2004}{\natexlab{b}})},\ \Eprint
  {http://arxiv.org/abs/astro-ph/0309686}{arXiv:astro-ph/0309686
  [astro-ph]}\BibitemShut {NoStop}%
\bibitem [{\citenamefont {Griest}\ and\ \citenamefont
  {Kamionkowski}(1990)}]{Griest:1989wd}%
  \BibitemOpen
  \bibfield  {author} {\bibinfo {author} {\bibfnamefont {K.}~\bibnamefont
  {Griest}}\ and\ \bibinfo {author} {\bibfnamefont {M.}~\bibnamefont
  {Kamionkowski}},\ }\href {\doibase 10.1103/PhysRevLett.64.615} {\bibfield
  {journal} {\bibinfo  {journal} {Phys. Rev. Lett.}\ }\textbf {\bibinfo
  {volume} {64}},\ \bibinfo {pages} {615} (\bibinfo {year} {1990})}\BibitemShut
  {NoStop}%
\bibitem [{\citenamefont {Hamaguchi}\ \emph {et~al.}(2007)\citenamefont
  {Hamaguchi}, \citenamefont {Shirai},\ and\ \citenamefont
  {Yanagida}}]{Hamaguchi:2007rb}%
  \BibitemOpen
  \bibfield  {author} {\bibinfo {author} {\bibfnamefont {K.}~\bibnamefont
  {Hamaguchi}}, \bibinfo {author} {\bibfnamefont {S.}~\bibnamefont {Shirai}}, \
  and\ \bibinfo {author} {\bibfnamefont {T.~T.}\ \bibnamefont {Yanagida}},\
  }\href {\doibase 10.1016/j.physletb.2007.08.047} {\bibfield  {journal}
  {\bibinfo  {journal} {Phys. Lett.}\ }\textbf {\bibinfo {volume} {B654}},\
  \bibinfo {pages} {110} (\bibinfo {year} {2007})},\ \Eprint
  {http://arxiv.org/abs/0707.2463}{arXiv:0707.2463 [hep-ph]}\BibitemShut
  {NoStop}%
\bibitem [{\citenamefont {Hamaguchi}\ \emph {et~al.}(2009)\citenamefont
  {Hamaguchi}, \citenamefont {Nakamura}, \citenamefont {Shirai},\ and\
  \citenamefont {Yanagida}}]{Hamaguchi:2008rv}%
  \BibitemOpen
  \bibfield  {author} {\bibinfo {author} {\bibfnamefont {K.}~\bibnamefont
  {Hamaguchi}}, \bibinfo {author} {\bibfnamefont {E.}~\bibnamefont {Nakamura}},
  \bibinfo {author} {\bibfnamefont {S.}~\bibnamefont {Shirai}}, \ and\ \bibinfo
  {author} {\bibfnamefont {T.~T.}\ \bibnamefont {Yanagida}},\ }\href {\doibase
  10.1016/j.physletb.2009.03.025} {\bibfield  {journal} {\bibinfo  {journal}
  {Phys. Lett.}\ }\textbf {\bibinfo {volume} {B674}},\ \bibinfo {pages} {299}
  (\bibinfo {year} {2009})},\ \Eprint
  {http://arxiv.org/abs/0811.0737}{arXiv:0811.0737 [hep-ph]}\BibitemShut
  {NoStop}%
\bibitem [{\citenamefont {Hamaguchi}\ \emph {et~al.}(2010)\citenamefont
  {Hamaguchi}, \citenamefont {Nakamura}, \citenamefont {Shirai},\ and\
  \citenamefont {Yanagida}}]{Hamaguchi:2009db}%
  \BibitemOpen
  \bibfield  {author} {\bibinfo {author} {\bibfnamefont {K.}~\bibnamefont
  {Hamaguchi}}, \bibinfo {author} {\bibfnamefont {E.}~\bibnamefont {Nakamura}},
  \bibinfo {author} {\bibfnamefont {S.}~\bibnamefont {Shirai}}, \ and\ \bibinfo
  {author} {\bibfnamefont {T.~T.}\ \bibnamefont {Yanagida}},\ }\href {\doibase
  10.1007/JHEP04(2010)119} {\bibfield  {journal} {\bibinfo  {journal} {JHEP}\
  }\textbf {\bibinfo {volume} {04}},\ \bibinfo {pages} {119} (\bibinfo {year}
  {2010})},\ \Eprint {http://arxiv.org/abs/0912.1683}{arXiv:0912.1683
  [hep-ph]}\BibitemShut {NoStop}%
\bibitem [{\citenamefont {Murayama}\ and\ \citenamefont
  {Shu}(2010)}]{Murayama:2009nj}%
  \BibitemOpen
  \bibfield  {author} {\bibinfo {author} {\bibfnamefont {H.}~\bibnamefont
  {Murayama}}\ and\ \bibinfo {author} {\bibfnamefont {J.}~\bibnamefont {Shu}},\
  }\href {\doibase 10.1016/j.physletb.2010.02.037} {\bibfield  {journal}
  {\bibinfo  {journal} {Phys. Lett.}\ }\textbf {\bibinfo {volume} {B686}},\
  \bibinfo {pages} {162} (\bibinfo {year} {2010})},\ \Eprint
  {http://arxiv.org/abs/0905.1720}{arXiv:0905.1720 [hep-ph]}\BibitemShut
  {NoStop}%
\bibitem [{\citenamefont {Hambye}\ and\ \citenamefont
  {Tytgat}(2010)}]{Hambye:2009fg}%
  \BibitemOpen
  \bibfield  {author} {\bibinfo {author} {\bibfnamefont {T.}~\bibnamefont
  {Hambye}}\ and\ \bibinfo {author} {\bibfnamefont {M.~H.~G.}\ \bibnamefont
  {Tytgat}},\ }\href {\doibase 10.1016/j.physletb.2009.11.050} {\bibfield
  {journal} {\bibinfo  {journal} {Phys. Lett.}\ }\textbf {\bibinfo {volume}
  {B683}},\ \bibinfo {pages} {39} (\bibinfo {year} {2010})},\ \Eprint
  {http://arxiv.org/abs/0907.1007}{arXiv:0907.1007 [hep-ph]}\BibitemShut
  {NoStop}%
\bibitem [{\citenamefont {Antipin}\ \emph
  {et~al.}(2015{\natexlab{a}})\citenamefont {Antipin}, \citenamefont {Redi},\
  and\ \citenamefont {Strumia}}]{Antipin:2014qva}%
  \BibitemOpen
  \bibfield  {author} {\bibinfo {author} {\bibfnamefont {O.}~\bibnamefont
  {Antipin}}, \bibinfo {author} {\bibfnamefont {M.}~\bibnamefont {Redi}}, \
  and\ \bibinfo {author} {\bibfnamefont {A.}~\bibnamefont {Strumia}},\ }\href
  {\doibase 10.1007/JHEP01(2015)157} {\bibfield  {journal} {\bibinfo  {journal}
  {JHEP}\ }\textbf {\bibinfo {volume} {01}},\ \bibinfo {pages} {157} (\bibinfo
  {year} {2015}{\natexlab{a}})},\ \Eprint
  {http://arxiv.org/abs/1410.1817}{arXiv:1410.1817 [hep-ph]}\BibitemShut
  {NoStop}%
\bibitem [{\citenamefont {Antipin}\ \emph
  {et~al.}(2015{\natexlab{b}})\citenamefont {Antipin}, \citenamefont {Redi},
  \citenamefont {Strumia},\ and\ \citenamefont {Vigiani}}]{Antipin:2015xia}%
  \BibitemOpen
  \bibfield  {author} {\bibinfo {author} {\bibfnamefont {O.}~\bibnamefont
  {Antipin}}, \bibinfo {author} {\bibfnamefont {M.}~\bibnamefont {Redi}},
  \bibinfo {author} {\bibfnamefont {A.}~\bibnamefont {Strumia}}, \ and\
  \bibinfo {author} {\bibfnamefont {E.}~\bibnamefont {Vigiani}},\ }\href
  {\doibase 10.1007/JHEP07(2015)039} {\bibfield  {journal} {\bibinfo  {journal}
  {JHEP}\ }\textbf {\bibinfo {volume} {07}},\ \bibinfo {pages} {039} (\bibinfo
  {year} {2015}{\natexlab{b}})},\ \Eprint
  {http://arxiv.org/abs/1503.08749}{arXiv:1503.08749 [hep-ph]}\BibitemShut
  {NoStop}%
\bibitem [{\citenamefont {Gross}\ \emph {et~al.}(2019)\citenamefont {Gross},
  \citenamefont {Mitridate}, \citenamefont {Redi}, \citenamefont {Smirnov},\
  and\ \citenamefont {Strumia}}]{Gross:2018zha}%
  \BibitemOpen
  \bibfield  {author} {\bibinfo {author} {\bibfnamefont {C.}~\bibnamefont
  {Gross}}, \bibinfo {author} {\bibfnamefont {A.}~\bibnamefont {Mitridate}},
  \bibinfo {author} {\bibfnamefont {M.}~\bibnamefont {Redi}}, \bibinfo {author}
  {\bibfnamefont {J.}~\bibnamefont {Smirnov}}, \ and\ \bibinfo {author}
  {\bibfnamefont {A.}~\bibnamefont {Strumia}},\ }\href {\doibase
  10.1103/PhysRevD.99.016024} {\bibfield  {journal} {\bibinfo  {journal} {Phys.
  Rev.}\ }\textbf {\bibinfo {volume} {D99}},\ \bibinfo {pages} {016024}
  (\bibinfo {year} {2019})},\ \Eprint
  {http://arxiv.org/abs/1811.08418}{arXiv:1811.08418 [hep-ph]}\BibitemShut
  {NoStop}%
\bibitem [{\citenamefont {Fukuda}\ \emph {et~al.}(2019)\citenamefont {Fukuda},
  \citenamefont {Luo},\ and\ \citenamefont {Shirai}}]{Fukuda:2018ufg}%
  \BibitemOpen
  \bibfield  {author} {\bibinfo {author} {\bibfnamefont {H.}~\bibnamefont
  {Fukuda}}, \bibinfo {author} {\bibfnamefont {F.}~\bibnamefont {Luo}}, \ and\
  \bibinfo {author} {\bibfnamefont {S.}~\bibnamefont {Shirai}},\ }\href
  {\doibase 10.1007/JHEP04(2019)107} {\bibfield  {journal} {\bibinfo  {journal}
  {JHEP}\ }\textbf {\bibinfo {volume} {04}},\ \bibinfo {pages} {107} (\bibinfo
  {year} {2019})},\ \Eprint {http://arxiv.org/abs/1812.02066}{arXiv:1812.02066
  [hep-ph]}\BibitemShut {NoStop}%
\bibitem [{\citenamefont {Bernstein}\ \emph {et~al.}(1985)\citenamefont
  {Bernstein}, \citenamefont {Brown},\ and\ \citenamefont
  {Feinberg}}]{Bernstein:1985th}%
  \BibitemOpen
  \bibfield  {author} {\bibinfo {author} {\bibfnamefont {J.}~\bibnamefont
  {Bernstein}}, \bibinfo {author} {\bibfnamefont {L.~S.}\ \bibnamefont
  {Brown}}, \ and\ \bibinfo {author} {\bibfnamefont {G.}~\bibnamefont
  {Feinberg}},\ }\href {\doibase 10.1103/PhysRevD.32.3261} {\bibfield
  {journal} {\bibinfo  {journal} {Phys. Rev.}\ }\textbf {\bibinfo {volume}
  {D32}},\ \bibinfo {pages} {3261} (\bibinfo {year} {1985})}\BibitemShut
  {NoStop}%
\bibitem [{\citenamefont {Srednicki}\ \emph {et~al.}(1988)\citenamefont
  {Srednicki}, \citenamefont {Watkins},\ and\ \citenamefont
  {Olive}}]{Srednicki:1988ce}%
  \BibitemOpen
  \bibfield  {author} {\bibinfo {author} {\bibfnamefont {M.}~\bibnamefont
  {Srednicki}}, \bibinfo {author} {\bibfnamefont {R.}~\bibnamefont {Watkins}},
  \ and\ \bibinfo {author} {\bibfnamefont {K.~A.}\ \bibnamefont {Olive}},\
  }\href {\doibase 10.1016/0550-3213(88)90099-5} {\bibfield  {journal}
  {\bibinfo  {journal} {Nucl. Phys.}\ }\textbf {\bibinfo {volume} {B310}},\
  \bibinfo {pages} {693} (\bibinfo {year} {1988})}\BibitemShut {NoStop}%
\bibitem [{\citenamefont {Griest}\ and\ \citenamefont
  {Seckel}(1991)}]{Griest:1990kh}%
  \BibitemOpen
  \bibfield  {author} {\bibinfo {author} {\bibfnamefont {K.}~\bibnamefont
  {Griest}}\ and\ \bibinfo {author} {\bibfnamefont {D.}~\bibnamefont
  {Seckel}},\ }\href {\doibase 10.1103/PhysRevD.43.3191} {\bibfield  {journal}
  {\bibinfo  {journal} {Phys. Rev.}\ }\textbf {\bibinfo {volume} {D43}},\
  \bibinfo {pages} {3191} (\bibinfo {year} {1991})}\BibitemShut {NoStop}%
\bibitem [{\citenamefont {Arcadi}\ \emph {et~al.}(2019)\citenamefont {Arcadi},
  \citenamefont {Djouadi},\ and\ \citenamefont {Raidal}}]{Arcadi:2019lka}%
  \BibitemOpen
  \bibfield  {author} {\bibinfo {author} {\bibfnamefont {G.}~\bibnamefont
  {Arcadi}}, \bibinfo {author} {\bibfnamefont {A.}~\bibnamefont {Djouadi}}, \
  and\ \bibinfo {author} {\bibfnamefont {M.}~\bibnamefont {Raidal}},\
  }\href@noop {} {\  (\bibinfo {year} {2019})},\ \Eprint
  {http://arxiv.org/abs/1903.03616}{arXiv:1903.03616 [hep-ph]}\BibitemShut
  {NoStop}%
\bibitem [{\citenamefont {Arcadi}\ \emph {et~al.}(2015)\citenamefont {Arcadi},
  \citenamefont {Mambrini},\ and\ \citenamefont {Richard}}]{Arcadi:2014lta}%
  \BibitemOpen
  \bibfield  {author} {\bibinfo {author} {\bibfnamefont {G.}~\bibnamefont
  {Arcadi}}, \bibinfo {author} {\bibfnamefont {Y.}~\bibnamefont {Mambrini}}, \
  and\ \bibinfo {author} {\bibfnamefont {F.}~\bibnamefont {Richard}},\ }\href
  {\doibase 10.1088/1475-7516/2015/03/018} {\bibfield  {journal} {\bibinfo
  {journal} {JCAP}\ }\textbf {\bibinfo {volume} {1503}},\ \bibinfo {pages}
  {018} (\bibinfo {year} {2015})},\ \Eprint
  {http://arxiv.org/abs/1411.2985}{arXiv:1411.2985 [hep-ph]}\BibitemShut
  {NoStop}%
\bibitem [{\citenamefont {Matsumoto}\ \emph {et~al.}(2014)\citenamefont
  {Matsumoto}, \citenamefont {Mukhopadhyay},\ and\ \citenamefont
  {Tsai}}]{Matsumoto:2014rxa}%
  \BibitemOpen
  \bibfield  {author} {\bibinfo {author} {\bibfnamefont {S.}~\bibnamefont
  {Matsumoto}}, \bibinfo {author} {\bibfnamefont {S.}~\bibnamefont
  {Mukhopadhyay}}, \ and\ \bibinfo {author} {\bibfnamefont {Y.-L.~S.}\
  \bibnamefont {Tsai}},\ }\href {\doibase 10.1007/JHEP10(2014)155} {\bibfield
  {journal} {\bibinfo  {journal} {JHEP}\ }\textbf {\bibinfo {volume} {10}},\
  \bibinfo {pages} {155} (\bibinfo {year} {2014})},\ \Eprint
  {http://arxiv.org/abs/1407.1859}{arXiv:1407.1859 [hep-ph]}\BibitemShut
  {NoStop}%
\bibitem [{\citenamefont {Cheung}\ \emph {et~al.}(2013)\citenamefont {Cheung},
  \citenamefont {Hall}, \citenamefont {Pinner},\ and\ \citenamefont
  {Ruderman}}]{Cheung:2012qy}%
  \BibitemOpen
  \bibfield  {author} {\bibinfo {author} {\bibfnamefont {C.}~\bibnamefont
  {Cheung}}, \bibinfo {author} {\bibfnamefont {L.~J.}\ \bibnamefont {Hall}},
  \bibinfo {author} {\bibfnamefont {D.}~\bibnamefont {Pinner}}, \ and\ \bibinfo
  {author} {\bibfnamefont {J.~T.}\ \bibnamefont {Ruderman}},\ }\href {\doibase
  10.1007/JHEP05(2013)100} {\bibfield  {journal} {\bibinfo  {journal} {JHEP}\
  }\textbf {\bibinfo {volume} {05}},\ \bibinfo {pages} {100} (\bibinfo {year}
  {2013})},\ \Eprint {http://arxiv.org/abs/1211.4873}{arXiv:1211.4873
  [hep-ph]}\BibitemShut {NoStop}%
\bibitem [{\citenamefont {Hamaguchi}\ and\ \citenamefont
  {Ishikawa}(2016)}]{Hamaguchi:2015rxa}%
  \BibitemOpen
  \bibfield  {author} {\bibinfo {author} {\bibfnamefont {K.}~\bibnamefont
  {Hamaguchi}}\ and\ \bibinfo {author} {\bibfnamefont {K.}~\bibnamefont
  {Ishikawa}},\ }\href {\doibase 10.1103/PhysRevD.93.055009} {\bibfield
  {journal} {\bibinfo  {journal} {Phys. Rev.}\ }\textbf {\bibinfo {volume}
  {D93}},\ \bibinfo {pages} {055009} (\bibinfo {year} {2016})},\ \Eprint
  {http://arxiv.org/abs/1510.05378}{arXiv:1510.05378 [hep-ph]}\BibitemShut
  {NoStop}%
\bibitem [{\citenamefont {Matsumoto}\ \emph {et~al.}(2016)\citenamefont
  {Matsumoto}, \citenamefont {Mukhopadhyay},\ and\ \citenamefont
  {Tsai}}]{Matsumoto:2016hbs}%
  \BibitemOpen
  \bibfield  {author} {\bibinfo {author} {\bibfnamefont {S.}~\bibnamefont
  {Matsumoto}}, \bibinfo {author} {\bibfnamefont {S.}~\bibnamefont
  {Mukhopadhyay}}, \ and\ \bibinfo {author} {\bibfnamefont {Y.-L.~S.}\
  \bibnamefont {Tsai}},\ }\href {\doibase 10.1103/PhysRevD.94.065034}
  {\bibfield  {journal} {\bibinfo  {journal} {Phys. Rev.}\ }\textbf {\bibinfo
  {volume} {D94}},\ \bibinfo {pages} {065034} (\bibinfo {year} {2016})},\
  \Eprint {http://arxiv.org/abs/1604.02230}{arXiv:1604.02230
  [hep-ph]}\BibitemShut {NoStop}%
\bibitem [{\citenamefont {Gondolo}\ and\ \citenamefont
  {Gelmini}(1991)}]{Gondolo:1990dk}%
  \BibitemOpen
  \bibfield  {author} {\bibinfo {author} {\bibfnamefont {P.}~\bibnamefont
  {Gondolo}}\ and\ \bibinfo {author} {\bibfnamefont {G.}~\bibnamefont
  {Gelmini}},\ }\href {\doibase 10.1016/0550-3213(91)90438-4} {\bibfield
  {journal} {\bibinfo  {journal} {Nucl. Phys.}\ }\textbf {\bibinfo {volume}
  {B360}},\ \bibinfo {pages} {145} (\bibinfo {year} {1991})}\BibitemShut
  {NoStop}%
\bibitem [{\citenamefont {Aghanim}\ \emph {et~al.}(2018)\citenamefont {Aghanim}
  \emph {et~al.}}]{Aghanim:2018eyx}%
  \BibitemOpen
  \bibfield  {author} {\bibinfo {author} {\bibfnamefont {N.}~\bibnamefont
  {Aghanim}} \emph {et~al.} (\bibinfo {collaboration} {Planck}),\ }\href@noop
  {} {\  (\bibinfo {year} {2018})},\ \Eprint
  {http://arxiv.org/abs/1807.06209}{arXiv:1807.06209 [astro-ph.CO]}\BibitemShut
  {NoStop}%
\bibitem [{\citenamefont {Saikawa}\ and\ \citenamefont
  {Shirai}(2018)}]{Saikawa:2018rcs}%
  \BibitemOpen
  \bibfield  {author} {\bibinfo {author} {\bibfnamefont {K.}~\bibnamefont
  {Saikawa}}\ and\ \bibinfo {author} {\bibfnamefont {S.}~\bibnamefont
  {Shirai}},\ }\href {\doibase 10.1088/1475-7516/2018/05/035} {\bibfield
  {journal} {\bibinfo  {journal} {JCAP}\ }\textbf {\bibinfo {volume} {1805}},\
  \bibinfo {pages} {035} (\bibinfo {year} {2018})},\ \Eprint
  {http://arxiv.org/abs/1803.01038}{arXiv:1803.01038 [hep-ph]}\BibitemShut
  {NoStop}%
\bibitem [{\citenamefont {Airapetian}\ \emph {et~al.}(2007)\citenamefont
  {Airapetian} \emph {et~al.}}]{Airapetian:2006vy}%
  \BibitemOpen
  \bibfield  {author} {\bibinfo {author} {\bibfnamefont {A.}~\bibnamefont
  {Airapetian}} \emph {et~al.} (\bibinfo {collaboration} {HERMES}),\ }\href
  {\doibase 10.1103/PhysRevD.75.012007} {\bibfield  {journal} {\bibinfo
  {journal} {Phys. Rev.}\ }\textbf {\bibinfo {volume} {D75}},\ \bibinfo {pages}
  {012007} (\bibinfo {year} {2007})},\ \Eprint
  {http://arxiv.org/abs/hep-ex/0609039}{arXiv:hep-ex/0609039
  [hep-ex]}\BibitemShut {NoStop}%
\bibitem [{\citenamefont {Bélanger}\ \emph {et~al.}(2018)\citenamefont
  {Bélanger}, \citenamefont {Boudjema}, \citenamefont {Goudelis},
  \citenamefont {Pukhov},\ and\ \citenamefont {Zaldivar}}]{Belanger:2018mqt}%
  \BibitemOpen
  \bibfield  {author} {\bibinfo {author} {\bibfnamefont {G.}~\bibnamefont
  {Bélanger}}, \bibinfo {author} {\bibfnamefont {F.}~\bibnamefont {Boudjema}},
  \bibinfo {author} {\bibfnamefont {A.}~\bibnamefont {Goudelis}}, \bibinfo
  {author} {\bibfnamefont {A.}~\bibnamefont {Pukhov}}, \ and\ \bibinfo {author}
  {\bibfnamefont {B.}~\bibnamefont {Zaldivar}},\ }\href {\doibase
  10.1016/j.cpc.2018.04.027} {\bibfield  {journal} {\bibinfo  {journal}
  {Comput. Phys. Commun.}\ }\textbf {\bibinfo {volume} {231}},\ \bibinfo
  {pages} {173} (\bibinfo {year} {2018})},\ \Eprint
  {http://arxiv.org/abs/1801.03509}{arXiv:1801.03509 [hep-ph]}\BibitemShut
  {NoStop}%
\bibitem [{\citenamefont {Aprile}\ \emph {et~al.}(2019)\citenamefont {Aprile}
  \emph {et~al.}}]{Aprile:2019dbj}%
  \BibitemOpen
  \bibfield  {author} {\bibinfo {author} {\bibfnamefont {E.}~\bibnamefont
  {Aprile}} \emph {et~al.},\ }\href@noop {} {\  (\bibinfo {year} {2019})},\
  \Eprint {http://arxiv.org/abs/1902.03234}{arXiv:1902.03234
  [astro-ph.CO]}\BibitemShut {NoStop}%
\bibitem [{\citenamefont {Amole}\ \emph {et~al.}(2019)\citenamefont {Amole}
  \emph {et~al.}}]{Amole:2019fdf}%
  \BibitemOpen
  \bibfield  {author} {\bibinfo {author} {\bibfnamefont {C.}~\bibnamefont
  {Amole}} \emph {et~al.} (\bibinfo {collaboration} {PICO}),\ }\href@noop {} {\
   (\bibinfo {year} {2019})},\ \Eprint
  {http://arxiv.org/abs/1902.04031}{arXiv:1902.04031 [astro-ph.CO]}\BibitemShut
  {NoStop}%
\bibitem [{\citenamefont {Mount}\ \emph {et~al.}(2017)\citenamefont {Mount}
  \emph {et~al.}}]{Mount:2017qzi}%
  \BibitemOpen
  \bibfield  {author} {\bibinfo {author} {\bibfnamefont {B.~J.}\ \bibnamefont
  {Mount}} \emph {et~al.},\ }\href@noop {} {\  (\bibinfo {year} {2017})},\
  \Eprint {http://arxiv.org/abs/1703.09144}{arXiv:1703.09144
  [physics.ins-det]}\BibitemShut {NoStop}%
\bibitem [{\citenamefont {Fallows}()}]{PICO500}%
  \BibitemOpen
  \bibfield  {author} {\bibinfo {author} {\bibfnamefont {S.}~\bibnamefont
  {Fallows}},\ }\href@noop {} {\bibinfo  {journal} {15th International
  Conference on Topics in Astroparticle and Underground Physics, TAUP2017
  \url{https://indico.cern.ch/event/606690/contributions/2623446}}\
  }\BibitemShut {NoStop}%
\bibitem [{\citenamefont {Ackermann}\ \emph {et~al.}(2015)\citenamefont
  {Ackermann} \emph {et~al.}}]{Ackermann:2015zua}%
  \BibitemOpen
\bibfield  {journal} {  }\bibfield  {author} {\bibinfo {author} {\bibfnamefont
  {M.}~\bibnamefont {Ackermann}} \emph {et~al.} (\bibinfo {collaboration}
  {Fermi-LAT}),\ }\href {\doibase 10.1103/PhysRevLett.115.231301} {\bibfield
  {journal} {\bibinfo  {journal} {Phys. Rev. Lett.}\ }\textbf {\bibinfo
  {volume} {115}},\ \bibinfo {pages} {231301} (\bibinfo {year} {2015})},\
  \Eprint {http://arxiv.org/abs/1503.02641}{arXiv:1503.02641
  [astro-ph.HE]}\BibitemShut {NoStop}%
\bibitem [{\citenamefont {Bartels}\ \emph {et~al.}(2012)\citenamefont
  {Bartels}, \citenamefont {Berggren},\ and\ \citenamefont
  {List}}]{Bartels:2012ex}%
  \BibitemOpen
  \bibfield  {author} {\bibinfo {author} {\bibfnamefont {C.}~\bibnamefont
  {Bartels}}, \bibinfo {author} {\bibfnamefont {M.}~\bibnamefont {Berggren}}, \
  and\ \bibinfo {author} {\bibfnamefont {J.}~\bibnamefont {List}},\ }\href
  {\doibase 10.1140/epjc/s10052-012-2213-9} {\bibfield  {journal} {\bibinfo
  {journal} {Eur. Phys. J.}\ }\textbf {\bibinfo {volume} {C72}},\ \bibinfo
  {pages} {2213} (\bibinfo {year} {2012})},\ \Eprint
  {http://arxiv.org/abs/1206.6639}{arXiv:1206.6639 [hep-ex]}\BibitemShut
  {NoStop}%
\bibitem [{\citenamefont {Patrignani}\ \emph {et~al.}(2016)\citenamefont
  {Patrignani} \emph {et~al.}}]{Patrignani:2016xqp}%
  \BibitemOpen
  \bibfield  {author} {\bibinfo {author} {\bibfnamefont {C.}~\bibnamefont
  {Patrignani}} \emph {et~al.} (\bibinfo {collaboration} {Particle Data
  Group}),\ }\href {\doibase 10.1088/1674-1137/40/10/100001} {\bibfield
  {journal} {\bibinfo  {journal} {Chin. Phys.}\ }\textbf {\bibinfo {volume}
  {C40}},\ \bibinfo {pages} {100001} (\bibinfo {year} {2016})}\BibitemShut
  {NoStop}%
\bibitem [{\citenamefont {Carena}\ \emph {et~al.}(2003)\citenamefont {Carena},
  \citenamefont {de~Gouvea}, \citenamefont {Freitas},\ and\ \citenamefont
  {Schmitt}}]{Carena:2003aj}%
  \BibitemOpen
  \bibfield  {author} {\bibinfo {author} {\bibfnamefont {M.}~\bibnamefont
  {Carena}}, \bibinfo {author} {\bibfnamefont {A.}~\bibnamefont {de~Gouvea}},
  \bibinfo {author} {\bibfnamefont {A.}~\bibnamefont {Freitas}}, \ and\
  \bibinfo {author} {\bibfnamefont {M.}~\bibnamefont {Schmitt}},\ }\href
  {\doibase 10.1103/PhysRevD.68.113007} {\bibfield  {journal} {\bibinfo
  {journal} {Phys. Rev.}\ }\textbf {\bibinfo {volume} {D68}},\ \bibinfo {pages}
  {113007} (\bibinfo {year} {2003})},\ \Eprint
  {http://arxiv.org/abs/hep-ph/0308053}{arXiv:hep-ph/0308053
  [hep-ph]}\BibitemShut {NoStop}%
\bibitem [{\citenamefont {Bicer}\ \emph {et~al.}(2014)\citenamefont {Bicer}
  \emph {et~al.}}]{Gomez-Ceballos:2013zzn}%
  \BibitemOpen
  \bibfield  {author} {\bibinfo {author} {\bibfnamefont {M.}~\bibnamefont
  {Bicer}} \emph {et~al.} (\bibinfo {collaboration} {TLEP Design Study Working
  Group}),\ }\bibfield  {booktitle} {\emph {\bibinfo {booktitle} {{Proceedings,
  2013 Community Summer Study on the Future of U.S. Particle Physics: Snowmass
  on the Mississippi (CSS2013): Minneapolis, MN, USA, July 29-August 6,
  2013}}},\ }\href {\doibase 10.1007/JHEP01(2014)164} {\bibfield  {journal}
  {\bibinfo  {journal} {JHEP}\ }\textbf {\bibinfo {volume} {01}},\ \bibinfo
  {pages} {164} (\bibinfo {year} {2014})},\ \Eprint
  {http://arxiv.org/abs/1308.6176}{arXiv:1308.6176 [hep-ex]}\BibitemShut
  {NoStop}%
\bibitem [{\citenamefont {Peskin}\ and\ \citenamefont
  {Takeuchi}(1990)}]{Peskin:1990zt}%
  \BibitemOpen
  \bibfield  {author} {\bibinfo {author} {\bibfnamefont {M.~E.}\ \bibnamefont
  {Peskin}}\ and\ \bibinfo {author} {\bibfnamefont {T.}~\bibnamefont
  {Takeuchi}},\ }\href {\doibase 10.1103/PhysRevLett.65.964} {\bibfield
  {journal} {\bibinfo  {journal} {Phys. Rev. Lett.}\ }\textbf {\bibinfo
  {volume} {65}},\ \bibinfo {pages} {964} (\bibinfo {year} {1990})}\BibitemShut
  {NoStop}%
\bibitem [{\citenamefont {Peskin}\ and\ \citenamefont
  {Takeuchi}(1992)}]{Peskin:1991sw}%
  \BibitemOpen
  \bibfield  {author} {\bibinfo {author} {\bibfnamefont {M.~E.}\ \bibnamefont
  {Peskin}}\ and\ \bibinfo {author} {\bibfnamefont {T.}~\bibnamefont
  {Takeuchi}},\ }\href {\doibase 10.1103/PhysRevD.46.381} {\bibfield  {journal}
  {\bibinfo  {journal} {Phys. Rev.}\ }\textbf {\bibinfo {volume} {D46}},\
  \bibinfo {pages} {381} (\bibinfo {year} {1992})}\BibitemShut {NoStop}%
\bibitem [{\citenamefont {Ciuchini}\ \emph {et~al.}(2013)\citenamefont
  {Ciuchini}, \citenamefont {Franco}, \citenamefont {Mishima},\ and\
  \citenamefont {Silvestrini}}]{Ciuchini:2013pca}%
  \BibitemOpen
  \bibfield  {author} {\bibinfo {author} {\bibfnamefont {M.}~\bibnamefont
  {Ciuchini}}, \bibinfo {author} {\bibfnamefont {E.}~\bibnamefont {Franco}},
  \bibinfo {author} {\bibfnamefont {S.}~\bibnamefont {Mishima}}, \ and\
  \bibinfo {author} {\bibfnamefont {L.}~\bibnamefont {Silvestrini}},\ }\href
  {\doibase 10.1007/JHEP08(2013)106} {\bibfield  {journal} {\bibinfo  {journal}
  {JHEP}\ }\textbf {\bibinfo {volume} {08}},\ \bibinfo {pages} {106} (\bibinfo
  {year} {2013})},\ \Eprint {http://arxiv.org/abs/1306.4644}{arXiv:1306.4644
  [hep-ph]}\BibitemShut {NoStop}%
\bibitem [{\citenamefont {de~Blas}\ \emph {et~al.}(2016)\citenamefont
  {de~Blas}, \citenamefont {Ciuchini}, \citenamefont {Franco}, \citenamefont
  {Mishima}, \citenamefont {Pierini}, \citenamefont {Reina},\ and\
  \citenamefont {Silvestrini}}]{deBlas:2016ojx}%
  \BibitemOpen
  \bibfield  {author} {\bibinfo {author} {\bibfnamefont {J.}~\bibnamefont
  {de~Blas}}, \bibinfo {author} {\bibfnamefont {M.}~\bibnamefont {Ciuchini}},
  \bibinfo {author} {\bibfnamefont {E.}~\bibnamefont {Franco}}, \bibinfo
  {author} {\bibfnamefont {S.}~\bibnamefont {Mishima}}, \bibinfo {author}
  {\bibfnamefont {M.}~\bibnamefont {Pierini}}, \bibinfo {author} {\bibfnamefont
  {L.}~\bibnamefont {Reina}}, \ and\ \bibinfo {author} {\bibfnamefont
  {L.}~\bibnamefont {Silvestrini}},\ }\href {\doibase 10.1007/JHEP12(2016)135}
  {\bibfield  {journal} {\bibinfo  {journal} {JHEP}\ }\textbf {\bibinfo
  {volume} {12}},\ \bibinfo {pages} {135} (\bibinfo {year} {2016})},\ \Eprint
  {http://arxiv.org/abs/1608.01509}{arXiv:1608.01509 [hep-ph]}\BibitemShut
  {NoStop}%
\bibitem [{\citenamefont {de~Blas}\ \emph {et~al.}(2017)\citenamefont
  {de~Blas}, \citenamefont {Ciuchini}, \citenamefont {Franco}, \citenamefont
  {Mishima}, \citenamefont {Pierini}, \citenamefont {Reina},\ and\
  \citenamefont {Silvestrini}}]{deBlas:2017wmn}%
  \BibitemOpen
  \bibfield  {author} {\bibinfo {author} {\bibfnamefont {J.}~\bibnamefont
  {de~Blas}}, \bibinfo {author} {\bibfnamefont {M.}~\bibnamefont {Ciuchini}},
  \bibinfo {author} {\bibfnamefont {E.}~\bibnamefont {Franco}}, \bibinfo
  {author} {\bibfnamefont {S.}~\bibnamefont {Mishima}}, \bibinfo {author}
  {\bibfnamefont {M.}~\bibnamefont {Pierini}}, \bibinfo {author} {\bibfnamefont
  {L.}~\bibnamefont {Reina}}, \ and\ \bibinfo {author} {\bibfnamefont
  {L.}~\bibnamefont {Silvestrini}},\ }\bibfield  {booktitle} {\emph {\bibinfo
  {booktitle} {{Proceedings, 2017 European Physical Society Conference on High
  Energy Physics (EPS-HEP 2017): Venice, Italy, July 5-12, 2017}}},\ }\href
  {\doibase 10.22323/1.314.0467} {\bibfield  {journal} {\bibinfo  {journal}
  {PoS}\ }\textbf {\bibinfo {volume} {EPS-HEP2017}},\ \bibinfo {pages} {467}
  (\bibinfo {year} {2017})},\ \Eprint
  {http://arxiv.org/abs/1710.05402}{arXiv:1710.05402 [hep-ph]}\BibitemShut
  {NoStop}%
\bibitem [{\citenamefont {Fan}\ \emph {et~al.}(2015)\citenamefont {Fan},
  \citenamefont {Reece},\ and\ \citenamefont {Wang}}]{Fan:2014vta}%
  \BibitemOpen
  \bibfield  {author} {\bibinfo {author} {\bibfnamefont {J.}~\bibnamefont
  {Fan}}, \bibinfo {author} {\bibfnamefont {M.}~\bibnamefont {Reece}}, \ and\
  \bibinfo {author} {\bibfnamefont {L.-T.}\ \bibnamefont {Wang}},\ }\href
  {\doibase 10.1007/JHEP09(2015)196} {\bibfield  {journal} {\bibinfo  {journal}
  {JHEP}\ }\textbf {\bibinfo {volume} {09}},\ \bibinfo {pages} {196} (\bibinfo
  {year} {2015})},\ \Eprint {http://arxiv.org/abs/1411.1054}{arXiv:1411.1054
  [hep-ph]}\BibitemShut {NoStop}%
\bibitem [{\citenamefont {Abdallah}\ \emph {et~al.}(2009)\citenamefont
  {Abdallah} \emph {et~al.}}]{Abdallah:2008aa}%
  \BibitemOpen
  \bibfield  {author} {\bibinfo {author} {\bibfnamefont {J.}~\bibnamefont
  {Abdallah}} \emph {et~al.} (\bibinfo {collaboration} {DELPHI}),\ }\href
  {\doibase 10.1140/epjc/s10052-009-0874-9} {\bibfield  {journal} {\bibinfo
  {journal} {Eur. Phys. J.}\ }\textbf {\bibinfo {volume} {C60}},\ \bibinfo
  {pages} {17} (\bibinfo {year} {2009})},\ \Eprint
  {http://arxiv.org/abs/0901.4486}{arXiv:0901.4486 [hep-ex]}\BibitemShut
  {NoStop}%
\bibitem [{\citenamefont {Alwall}\ \emph {et~al.}(2014)\citenamefont {Alwall},
  \citenamefont {Frederix}, \citenamefont {Frixione}, \citenamefont {Hirschi},
  \citenamefont {Maltoni}, \citenamefont {Mattelaer}, \citenamefont {Shao},
  \citenamefont {Stelzer}, \citenamefont {Torrielli},\ and\ \citenamefont
  {Zaro}}]{Alwall:2014hca}%
  \BibitemOpen
  \bibfield  {author} {\bibinfo {author} {\bibfnamefont {J.}~\bibnamefont
  {Alwall}}, \bibinfo {author} {\bibfnamefont {R.}~\bibnamefont {Frederix}},
  \bibinfo {author} {\bibfnamefont {S.}~\bibnamefont {Frixione}}, \bibinfo
  {author} {\bibfnamefont {V.}~\bibnamefont {Hirschi}}, \bibinfo {author}
  {\bibfnamefont {F.}~\bibnamefont {Maltoni}}, \bibinfo {author} {\bibfnamefont
  {O.}~\bibnamefont {Mattelaer}}, \bibinfo {author} {\bibfnamefont {H.~S.}\
  \bibnamefont {Shao}}, \bibinfo {author} {\bibfnamefont {T.}~\bibnamefont
  {Stelzer}}, \bibinfo {author} {\bibfnamefont {P.}~\bibnamefont {Torrielli}},
  \ and\ \bibinfo {author} {\bibfnamefont {M.}~\bibnamefont {Zaro}},\ }\href
  {\doibase 10.1007/JHEP07(2014)079} {\bibfield  {journal} {\bibinfo  {journal}
  {JHEP}\ }\textbf {\bibinfo {volume} {07}},\ \bibinfo {pages} {079} (\bibinfo
  {year} {2014})},\ \Eprint {http://arxiv.org/abs/1405.0301}{arXiv:1405.0301
  [hep-ph]}\BibitemShut {NoStop}%
\bibitem [{\citenamefont {Sjostrand}\ \emph {et~al.}(2008)\citenamefont
  {Sjostrand}, \citenamefont {Mrenna},\ and\ \citenamefont
  {Skands}}]{Sjostrand:2007gs}%
  \BibitemOpen
  \bibfield  {author} {\bibinfo {author} {\bibfnamefont {T.}~\bibnamefont
  {Sjostrand}}, \bibinfo {author} {\bibfnamefont {S.}~\bibnamefont {Mrenna}}, \
  and\ \bibinfo {author} {\bibfnamefont {P.~Z.}\ \bibnamefont {Skands}},\
  }\href {\doibase 10.1016/j.cpc.2008.01.036} {\bibfield  {journal} {\bibinfo
  {journal} {Comput. Phys. Commun.}\ }\textbf {\bibinfo {volume} {178}},\
  \bibinfo {pages} {852} (\bibinfo {year} {2008})},\ \Eprint
  {http://arxiv.org/abs/0710.3820}{arXiv:0710.3820 [hep-ph]}\BibitemShut
  {NoStop}%
\bibitem [{\citenamefont {Fox}\ \emph {et~al.}(2011)\citenamefont {Fox},
  \citenamefont {Harnik}, \citenamefont {Kopp},\ and\ \citenamefont
  {Tsai}}]{Fox:2011fx}%
  \BibitemOpen
  \bibfield  {author} {\bibinfo {author} {\bibfnamefont {P.~J.}\ \bibnamefont
  {Fox}}, \bibinfo {author} {\bibfnamefont {R.}~\bibnamefont {Harnik}},
  \bibinfo {author} {\bibfnamefont {J.}~\bibnamefont {Kopp}}, \ and\ \bibinfo
  {author} {\bibfnamefont {Y.}~\bibnamefont {Tsai}},\ }\href {\doibase
  10.1103/PhysRevD.84.014028} {\bibfield  {journal} {\bibinfo  {journal} {Phys.
  Rev.}\ }\textbf {\bibinfo {volume} {D84}},\ \bibinfo {pages} {014028}
  (\bibinfo {year} {2011})},\ \Eprint
  {http://arxiv.org/abs/1103.0240}{arXiv:1103.0240 [hep-ph]}\BibitemShut
  {NoStop}%
\bibitem [{\citenamefont {Aaboud}\ \emph {et~al.}(2017)\citenamefont {Aaboud}
  \emph {et~al.}}]{Aaboud:2017dor}%
  \BibitemOpen
  \bibfield  {author} {\bibinfo {author} {\bibfnamefont {M.}~\bibnamefont
  {Aaboud}} \emph {et~al.} (\bibinfo {collaboration} {ATLAS}),\ }\href
  {\doibase 10.1140/epjc/s10052-017-4965-8} {\bibfield  {journal} {\bibinfo
  {journal} {Eur. Phys. J.}\ }\textbf {\bibinfo {volume} {C77}},\ \bibinfo
  {pages} {393} (\bibinfo {year} {2017})},\ \Eprint
  {http://arxiv.org/abs/1704.03848}{arXiv:1704.03848 [hep-ex]}\BibitemShut
  {NoStop}%
\bibitem [{\citenamefont {Aaboud}\ \emph {et~al.}(2018)\citenamefont {Aaboud}
  \emph {et~al.}}]{Aaboud:2017phn}%
  \BibitemOpen
  \bibfield  {author} {\bibinfo {author} {\bibfnamefont {M.}~\bibnamefont
  {Aaboud}} \emph {et~al.} (\bibinfo {collaboration} {ATLAS}),\ }\href
  {\doibase 10.1007/JHEP01(2018)126} {\bibfield  {journal} {\bibinfo  {journal}
  {JHEP}\ }\textbf {\bibinfo {volume} {01}},\ \bibinfo {pages} {126} (\bibinfo
  {year} {2018})},\ \Eprint {http://arxiv.org/abs/1711.03301}{arXiv:1711.03301
  [hep-ex]}\BibitemShut {NoStop}%
\bibitem [{\citenamefont {Ellis}\ \emph {et~al.}(2018)\citenamefont {Ellis},
  \citenamefont {Fowlie}, \citenamefont {Marzola},\ and\ \citenamefont
  {Raidal}}]{Balazs:2017ple}%
  \BibitemOpen
  \bibfield  {author} {\bibinfo {author} {\bibfnamefont {J.}~\bibnamefont
  {Ellis}}, \bibinfo {author} {\bibfnamefont {A.}~\bibnamefont {Fowlie}},
  \bibinfo {author} {\bibfnamefont {L.}~\bibnamefont {Marzola}}, \ and\
  \bibinfo {author} {\bibfnamefont {M.}~\bibnamefont {Raidal}},\ }\href
  {\doibase 10.1103/PhysRevD.97.115014} {\bibfield  {journal} {\bibinfo
  {journal} {Phys. Rev.}\ }\textbf {\bibinfo {volume} {D97}},\ \bibinfo {pages}
  {115014} (\bibinfo {year} {2018})},\ \Eprint
  {http://arxiv.org/abs/1711.09912}{arXiv:1711.09912 [hep-ph]}\BibitemShut
  {NoStop}%
\bibitem [{\citenamefont {Chen}\ and\ \citenamefont
  {Yokoya}(1988)}]{Chen:1988pn}%
  \BibitemOpen
  \bibfield  {author} {\bibinfo {author} {\bibfnamefont {P.}~\bibnamefont
  {Chen}}\ and\ \bibinfo {author} {\bibfnamefont {K.}~\bibnamefont {Yokoya}},\
  }\href {\doibase 10.1103/PhysRevD.38.987} {\bibfield  {journal} {\bibinfo
  {journal} {Phys. Rev.}\ }\textbf {\bibinfo {volume} {D38}},\ \bibinfo {pages}
  {987} (\bibinfo {year} {1988})}\BibitemShut {NoStop}%
\bibitem [{\citenamefont {Peskin}(1999)}]{Peskin:1999pk}%
  \BibitemOpen
  \bibfield  {author} {\bibinfo {author} {\bibfnamefont {M.~E.}\ \bibnamefont
  {Peskin}},\ }\href@noop {} {\bibfield  {journal} {\bibinfo  {journal}
  {SLAC-TN-04-032, LCC-0010}\ } (\bibinfo {year} {1999})}\BibitemShut {NoStop}%
\bibitem [{\citenamefont {Datta}\ \emph {et~al.}(2005)\citenamefont {Datta},
  \citenamefont {Kong},\ and\ \citenamefont {Matchev}}]{Datta:2005gm}%
  \BibitemOpen
  \bibfield  {author} {\bibinfo {author} {\bibfnamefont {A.~K.}\ \bibnamefont
  {Datta}}, \bibinfo {author} {\bibfnamefont {K.}~\bibnamefont {Kong}}, \ and\
  \bibinfo {author} {\bibfnamefont {K.~T.}\ \bibnamefont {Matchev}},\
  }\bibfield  {booktitle} {\emph {\bibinfo {booktitle} {{2005 International
  Linear Collider Workshop : LCWS 2005 : Stanford, California, USA, 18-22
  March, 2005}}},\ }\href@noop {} {\bibfield  {journal} {\bibinfo  {journal}
  {eConf}\ }\textbf {\bibinfo {volume} {C050318}},\ \bibinfo {pages} {0215}
  (\bibinfo {year} {2005})},\ \Eprint
  {http://arxiv.org/abs/hep-ph/0508161}{arXiv:hep-ph/0508161
  [hep-ph]}\BibitemShut {NoStop}%
\bibitem [{\citenamefont {Kuraev}\ and\ \citenamefont
  {Fadin}(1985)}]{Kuraev:1985hb}%
  \BibitemOpen
  \bibfield  {author} {\bibinfo {author} {\bibfnamefont {E.~A.}\ \bibnamefont
  {Kuraev}}\ and\ \bibinfo {author} {\bibfnamefont {V.~S.}\ \bibnamefont
  {Fadin}},\ }\href@noop {} {\bibfield  {journal} {\bibinfo  {journal} {Sov. J.
  Nucl. Phys.}\ }\textbf {\bibinfo {volume} {41}},\ \bibinfo {pages} {466}
  (\bibinfo {year} {1985})},\ \bibinfo {note} {[Yad.
  Fiz.41,733(1985)]}\BibitemShut {NoStop}%
\bibitem [{\citenamefont {Adloff}\ \emph {et~al.}(2009)\citenamefont {Adloff}
  \emph {et~al.}}]{Adloff:2008aa}%
  \BibitemOpen
  \bibfield  {author} {\bibinfo {author} {\bibfnamefont {C.}~\bibnamefont
  {Adloff}} \emph {et~al.} (\bibinfo {collaboration} {CALICE}),\ }\href
  {\doibase 10.1016/j.nima.2009.07.026} {\bibfield  {journal} {\bibinfo
  {journal} {Nucl. Instrum. Meth.}\ }\textbf {\bibinfo {volume} {A608}},\
  \bibinfo {pages} {372} (\bibinfo {year} {2009})},\ \Eprint
  {http://arxiv.org/abs/0811.2354}{arXiv:0811.2354
  [physics.ins-det]}\BibitemShut {NoStop}%
\end{thebibliography}%


%

\end{document}